\documentclass[fleqn,usenatbib]{mnras}

\usepackage{newtxtext,newtxmath}

\usepackage[T1]{fontenc}
\usepackage{ae,aecompl}

\usepackage{graphicx}	
\usepackage{amsmath}	
\usepackage{hyperref}
\usepackage{natbib}
\usepackage[caption=false]{subfig}
\usepackage{ragged2e}
\usepackage{placeins}
\usepackage{bigints} 
\usepackage{physics}
\usepackage{xcolor}
\usepackage{cleveref}
\usepackage[export]{adjustbox}
\usepackage{longtable}
\usepackage{xtab}
\usepackage{bm}

\usepackage{scalerel,tikz}
\usetikzlibrary{svg.path}
\definecolor{orcidlogocol}{HTML}{A6CE39}
\tikzset{orcidlogo/.pic={
 \fill[orcidlogocol] svg{M256,128c0,70.7-57.3,128-128,128C57.3,256,0,198.7,0,128C0,57.3,57.3,0,128,0C198.7,0,256,57.3,256,128z};
 \fill[white] svg{M86.3,186.2H70.9V79.1h15.4v48.4V186.2z}
 svg{M108.9,79.1h41.6c39.6,0,57,28.3,57,53.6c0,27.5-21.5,53.6-56.8,53.6h-41.8V79.1z M124.3,172.4h24.5c34.9,0,42.9-26.5,42.9-39.7c0-21.5-13.7-39.7-43.7-39.7h-23.7V172.4z}
 svg{M88.7,56.8c0,5.5-4.5,10.1-10.1,10.1c-5.6,0-10.1-4.6-10.1-10.1c0-5.6,4.5-10.1,10.1-10.1C84.2,46.7,88.7,51.3,88.7,56.8z};
}}
\newcommand\orcidicon[1]{\href{https://orcid.org/#1}{\mbox{\scalerel*{
\begin{tikzpicture}[yscale=-1,transform shape]
\pic{orcidlogo};
\end{tikzpicture}
}{|}}}}

\def\mean#1{\left< #1 \right>}
\def\sun{\odot}

\defcitealias{Mohapatra2022characterising}{M22b}
\defcitealias{Mohapatra2022VSF}{M22a}

\title[Cooling vs.~buoyancy vs.~mixing]{Multiphase condensation in cluster halos: interplay of cooling, buoyancy and mixing}

\author[Mohapatra et al]{
Rajsekhar Mohapatra$^{\orcidicon{0000-0002-1600-7552}\,1,2}$\thanks{E-mail: rmohapatra@princeton.edu (RM)},
Prateek Sharma$^{\orcidicon{0000-0003-2635-4643}\,3}$\thanks{E-mail: prateek@iisc.ac.in (PS)},
Christoph Federrath$^{\orcidicon{0000-0002-0706-2306}\,2,4}$\thanks{E-mail: christoph.federrath@anu.edu.au (CF)} and \newauthor
Eliot Quataert$^{\orcidicon{0000-0001-9185-5044}\,1}$\thanks{E-mail: quataert@princeton.edu (EQ)}
\\
$^{1}$Department of Astrophysical Sciences, Princeton University, Princeton, NJ 08544, USA\\
$^{2}$Research School of Astronomy and Astrophysics, Australian National University, Canberra, ACT~2611, Australia\\
$^{3}$Department of Physics, Indian Institute of Science, Bangalore, KA 560012, India\\
$^{4}$Australian Research Council Centre of Excellence in All Sky Astrophysics (ASTRO3D), Canberra, ACT~2611, Australia
}

\date{Accepted XXX. Received YYY; in original form ZZZ}

\pubyear{2022}

\begin{document}
\label{firstpage}
\pagerange{\pageref{firstpage}--\pageref{lastpage}}
\maketitle

\begin{abstract}
Gas in the central regions of cool-core clusters and other massive halos has a short cooling time ($\lesssim1~\mathrm{Gyr}$). Theoretical models predict that this gas is susceptible to multiphase condensation, in which cold gas is expected to condense out of the hot phase if the ratio of the thermal instability growth time scale ($t_{\mathrm{ti}}$) to the free-fall time ($t_{\mathrm{ff}}$) is $t_{\mathrm{ti}}/t_{\mathrm{ff}}\lesssim10$. The turbulent mixing time $t_{\mathrm{mix}}$ is another important time scale: if $t_{\mathrm{mix}}$ is short enough, the fluctuations are mixed before they can cool.
In this study, we perform high-resolution ($512^2\times768$--$1024^2\times1536$ resolution elements) hydrodynamic simulations of turbulence in a stratified medium, including radiative cooling of the gas. We explore the parameter space of $t_{\mathrm{ti}}/t_{\mathrm{ff}}$ and $t_{\mathrm{ti}}/t_{\mathrm{mix}}$ relevant to galaxy and cluster halos. We also study the effect of the steepness of the entropy profile, the strength of turbulent forcing and the nature of turbulent forcing (natural mixture vs.~compressive modes) on 
multiphase gas condensation.  
We find that larger values of $t_{\mathrm{ti}}/t_{\mathrm{ff}}$ or $t_{\mathrm{ti}}/t_{\mathrm{mix}}$ generally imply stability against multiphase gas condensation, whereas larger density fluctuations (e.g., due to compressible turbulence) promote multiphase gas condensation. We propose a new criterion $\min(t_{\mathrm{ti}}/\min(t_{\mathrm{mix}},t_\mathrm{ff}))\lesssim c_2\times\exp(c_1\sigma_s)$ for when the halo becomes multiphase, where $\sigma_s$ denotes the amplitude of logarithmic density fluctuations and $c_1\simeq6$, $c_2\simeq1.8$ from an empirical fit to our results.
\end{abstract}




\section{Introduction}\label{sec:introduction}
Galaxy clusters are the largest gravitationally relaxed objects in the universe. Based on the central temperature/entropy of the gas in their central regions, clusters are broadly divided into two types--cool cores (CC) and non-cool cores (NCC). CC clusters cool radiatively and in the absence of any external heating, they can generate massive cooling flows ($100$--$1000~\mathrm{M}_\sun/\mathrm{yr}$) \citep{fabian1994cooling}. Such massive cooling flows are not observed in most clusters and the brightest cluster galaxies (BCGs) are rarely star-forming. 
Heating by energy injected from the active galactic nucleus (AGN) is expected to offset the cooling in galaxy clusters---the net mechanical energy input from the AGN, estimated from X-ray cavities roughly balances out the cooling \citep{Fabian2012ARA&A,McNamara2012NJPh,Olivares2022MNRAS}. 

While the ICM is expected to be in global thermal balance, localised density perturbations can lead to condensation of cold gas from the hot medium. Filaments of atomic gas (at $\sim10^4~\mathrm{K}$) and molecular gas (at $\sim10~\mathrm{K}$) are seen ubiquitously, often co-spatial with dense regions in the hotter ($10^7$--$10^8~\mathrm{K}$) X-ray emitting phase \citep{Werner2013ApJ,Anderson2018A&A,Olivares2019A&A}. 
Theoretical studies such as \cite{sharma2012thermal,mccourt2012,Voit2017ApJ} point towards the existence of a critical value of the ratio between the hot gas cooling time ($t_{\mathrm{cool}}$) and the free-fall time ($t_{\mathrm{ff}}$), i.e., $t_{\mathrm{cool}}/t_{\mathrm{ff}}$. If $t_{\mathrm{cool}}/t_{\mathrm{ff}}\lesssim10$, then seed perturbations in the thermally unstable hot gas lead to the condensation of cold gas. Multi-wavelength observations of clusters also show the existence of cold gas in cluster cores around regions where $t_{\mathrm{cool}}/t_{\mathrm{ff}}\lesssim10$--$20$ \citep{Voit2015ApJ,Lakhchaura2018MNRAS,Olivares2019A&A,OSullivan2021MNRAS}. 

Numerical simulations offer us some further insights. Cluster-scale simulations including AGN feedback loop such as \cite{prasad2015,Beckmann2019A&A} show that galaxy clusters go through cycles of gas condensation (when $t_{\mathrm{cool}/t_{\mathrm{ff}}}\lesssim10$). Mass accretion onto the central super-massive black hole (SMBH), which releases jets that heat the ICM, raises the value of $t_{\mathrm{cool}}/t_{\mathrm{ff}}$ to prevent further condensation. Once the heating stops due to a lack of mass accretion, cooling takes over and this cycle repeats. 

However, there are some challenges to these models. \citet{Choudhury2019} show that the threshold for cluster atmospheres to be thermally stable increases with increasing amplitude of seed density fluctuations. \cite{Nelson2020MNRAS} study the formation of small-scale cold gas in the circumgalactic medium (CGM) of galaxies in the TNG50 simulations. They find that cold clouds form due to large (order unity) perturbations in the gas density, which can trigger multiphase condensation in halos with $t_{\mathrm{cool}}/t_{\mathrm{ff}}>10$. \cite{Choudhury2019} show that the threshold condition for multiphase condensation applies to the local value of $t_{\mathrm{cool}}/t_{\mathrm{ff}}$, rather than its globally-averaged value. On a similar note, \cite{Voit2021ApJ} proposes that locally $t_{\mathrm{cool}}/t_{\mathrm{ff}}\lesssim1$ leads to condensation but on a global scale the threshold condition depends on the amplitude of entropy fluctuations.

Turbulence plays a critical role in the evolution of the ICM. It is driven on large scales ($\sim100$--$500~\mathrm{kpc}$) by galaxy motions during mergers and on smaller scales by AGN ($\sim10$--$100~\mathrm{kpc}$). It can transfer the heat from the gas heated by AGN jets to the ambient ICM through turbulent mixing \citep{banerjee2014turbulence} and viscous dissipation. Further, \cite{Voit2018ApJ} shows that turbulence can drive buoyancy oscillations that lead  to condensation when $10\lesssim t_{\mathrm{cool}}/t_{\mathrm{ff}}\lesssim20$.
\cite{Gaspari2018ApJ} argue that the turbulent mixing time $t_{\mathrm{mix}}$ is a more important timescale than $t_{\mathrm{ff}}$, and the regions with cold gas are traced better by $t_{\mathrm{cool}}/t_{\mathrm{mix}}\lesssim1$. \cite{Mohapatra2019} show that the onset of multiphase condensation is delayed when one drives turbulence on smaller scales, since $t_{\mathrm{mix}}$ is shorter for small-scale driving. 

\cite{Olivares2019A&A,OSullivan2021MNRAS} find that $t_{\mathrm{cool}}/t_{\mathrm{mix}}\approx1$ in regions of clusters where the cold-phase gas is observed. 
However, it is difficult to disentangle the importance of the two ratios ($t_{\mathrm{cool}}/t_{\mathrm{ff}})$ and $t_{\mathrm{cool}}/t_{\mathrm{mix}}$) from observations, since (1) $t_{\mathrm{cool}}$ varies more strongly with radius compared to $t_{\mathrm{ff}}$ and $t_{\mathrm{mix}}$ in cluster centers, and (2) we do not have many direct observations of turbulent velocities of the hot phase, except by Hitomi for the Perseus cluster \citep{hitomi2016}. Hence we rely on indirect methods of constraining turbulence and $t_{\mathrm{mix}}$ \citep[see][for a review]{Simionescu2019SSRv}.  

Turbulence plays a dual role in multiphase condensation. On one hand, turbulence drives large density fluctuations on the driving scale in the ICM, leading to multiphase gas condensation. On the other hand, turbulent mixing suppresses the density contrast and multiphase condensation. \cite{Baek2022ApJ} find molecular gas co-spatial with sloshing features seen in the X-ray emission, implying that the velocity field affects condensation locally. Using idealised simulations, \cite{Mohapatra2020,Mohapatra2021MNRAS,Mohapatra2022MNRASc} have shown that the amplitude of 
turbulence-driven (other sources, e.g., cooling, buoyancy, jet/outflows can also drive density fluctuations) density fluctuations depends on the degree of stratification of the ICM, the turbulent Mach number and the nature of driving (solenoidal vs.~compressive modes). However, many previous theoretical and numerical studies of the ICM initialise seed density fluctuations by hand, independent of the gas turbulence.

In order to better constrain the conditions required for the onset of multiphase condensation and to separate the two proposed threshold ratios of the time-scales, we conduct high-resolution hydrodynamic simulations of turbulence in a stratified medium, including radiative cooling of the gas. In our study, density fluctuations develop naturally due to the large-scale turbulence driving. We vary four main parameters relevant to cluster halos---(1) the strength of stratification, which controls $t_{\mathrm{ff}}$, (2) the strength and (3) the nature of turbulence forcing, which controls $t_{\mathrm{mix}}$ and the amplitude of density fluctuations, and (4) the initial gas density, which controls $t_{\mathrm{cool}}$.

This paper is organised as follows. We introduce our model, numerical setup and tools in \cref{sec:Methods}. Then we present our results and discuss them in the context of galaxy cluster halos in \cref{sec:results-discussion}. We summarise our key findings regarding the two timescale ratios in \cref{sec:timescale_ratio_summary}. In \cref{sec:caveats-future}, we discuss some of the shortcomings of our model and setup, missing physics and how they might affect our results as well as the future prospects of this work. Finally, we present our concluding remarks in \cref{sec:Conclusion}.

\section{Methods}\label{sec:Methods}
\subsection{Model equations}\label{subsec:ModEq}
We use Euler equations to model the ICM, with acceleration due to gravity ($\bm{g}$) and turbulence ($\bm{a}$), radiative cooling with a rate density $\mathcal{L}$, and thermal heating with a rate density $Q$ as additional source terms. We assume an ideal gas equation of state with an adiabatic index $\gamma=5/3$. We evolve the following equations:
\begin{subequations}
	\begin{align}
	\label{eq:continuity}
	&\frac{\partial\rho}{\partial t}+\nabla\cdot (\rho \bm{v})=0,\\
	\label{eq:momentum}
	&\frac{\partial(\rho\bm{v})}{\partial t}+\nabla\cdot (\rho\bm{v}\otimes\bm{v})+ \nabla P=\rho(\bm{a}+ \bm{g}),\\
	\label{eq:energy}
	&\frac{\partial E}{\partial t}+\nabla\cdot ((E+P)\bm{v})=\rho \bm{v}\cdot(\bm{a}+\bm{g})+Q-\mathcal{L},\\
	\label{eq:tot_energy}
	&E=\frac{\rho\bm{v}\cdot\bm{v}}{2} + \frac{P}{\gamma-1},
	\end{align}
\end{subequations}
where $\rho$ is the gas mass density, $\bm{v}$ is the velocity, $P=\rho k_B T/(\mu m_p)$ is the thermal pressure, $\mu$ is the mean particle weight, $m_p$ is the proton mass, $k_B$ is the Boltzmann constant, and $T$ is the temperature. In the energy equation (eq.~\ref{eq:energy}), the total energy density is given by $E$ and the cooling rate density $\mathcal{L}$ is given by
\begin{equation}\label{eq:cooling_function}
\mathcal{L} = n_en_i\Lambda(T),
\end{equation}
where $n_e$ and $n_i$ 
are the electron and ion number densities, respectively. We use the temperature-dependent cooling function $\Lambda(T)$.  

\subsection{Important timescales}\label{subsec:imp_timescales}
The timescales of interest in this study are---the gas cooling time $t_{\mathrm{cool}}$, the isobaric thermal instability growth time $t_{\mathrm{ti}}$, the sound crossing time $t_{\mathrm{cs}}$, the gas freefall time $t_{\mathrm{ff}}$ and the turbulent mixing time on the driving scale $t_{\mathrm{mix}}$. They are defined as follows:
\begin{subequations}
    \begin{align}
        &t_{\mathrm{cool}}=\frac{P}{(\gamma-1)\mathcal{L}}\text{,} \label{eq:t_cool}\\
        &t_{\mathrm{ti}} = \frac{\gamma t_{\mathrm{cool}}}{2-\mathrm{d}\ln{\Lambda(T)/\mathrm{d}\ln{T}-\alpha}},\label{eq:t_ti}\\
        &t_{\mathrm{cs}} =\frac{L}{c_\mathrm{s}}\text{,} \label{eq:t_cs} \\
        &t_{\mathrm{ff}} =\sqrt{\frac{2L}{g}}\text{ and} \label{eq:t_ff} \\
        &t_{\mathrm{mix}}=\frac{\ell_{\mathrm{driv}}}{v_{\mathrm{\ell_{\mathrm{driv}}}}}\simeq \frac{L}{2v}\label{eq:t_mix}\text{,}
    \end{align}
\end{subequations}
where $\alpha$ characterises the density dependence of the heating rate density $Q$, with $Q\propto\rho^\alpha$. The sound speed $c_\mathrm{s}$ is given by $\sqrt{\gamma P/\rho}$. For a derivation of \cref{eq:t_ti} using linear stability analysis, see section 4.1 in \cite{mccourt2012}. The two scales $L$ and $\ell_\mathrm{driv}$ denote the size of the system and the driving scale of turbulence, respectively. In our simulations, $\ell_{\mathrm{driv}}=L/2$ and $v_{\ell_{\mathrm{driv}}}\approx v$, so $t_{\mathrm{mix}}\simeq L/(2v)$.

\subsection{Numerical methods}\label{subsec:numerical_methods}
We use a modified version of the FLASH code \citep{Fryxell2000,Dubey2008}, version~4, to solve \cref{eq:continuity} to \cref{eq:tot_energy} in our simulations. For time integration, we use the MUSCL-Hancock scheme \citep{van1984SIAM,Waagan2009JCoPh} with the HLL5R approximate Riemann scheme \citep{Waagan2011}. We use a second-order reconstruction method that uses primitive variables and ensures that density and internal energy are positive. Our simulation domain size is the same as in \cite{Mohapatra2020}---we use a cuboidal box with $L_x=L_y=L=40~\mathrm{kpc}$ and $L_z=1.5L=60~\mathrm{kpc}$. The box is centred at the origin $(0,0,0)$. We implement periodic boundary conditions along the $x$ and $y$ direction for all variables. In the $z$ direction, we implement diode boundary conditions for the velocity. For density and pressure, we fix the values in the guard cells to their initial values throughout the duration of the simulation. In addition to using a larger box along the $z$ direction to minimise the effect of the boundaries, we further smoothly decay the source terms---turbulent acceleration $\bm{a}$, gas cooling rate density $\mathcal{L}$, and gas heating rate $Q$ for $|z|>L/2$, where the weighting function $w(z)$ is given by
\begin{align}
    w(z)&=1 \text{ for }|z|/L\leq0.5 \text{ and}\nonumber\\
       &=\exp(-((2|z|/L-1)/0.15)^2) \text{ for }|z|/L>0.5.\label{z-weighting function}
\end{align}
We analyse the outputs from our simulations only in the central cubical region with $|x|$, $|y|$, $|z|<L/2$.

\subsection{Problem setup}\label{subsec:setup}
\subsubsection{Initial density and pressure profiles} \label{subsubsec:init_dens_pres_profiles}
We set up a gravitationally stratified atmosphere with a constant $\bm{g}$  oriented along the $-\hat{\bm{z}}$ direction. Pressure and density follow exponential profiles along the $z$ direction at time $t=0$ and the gas is at hydrostatic equilibrium,  given by
\begin{subequations}
\begin{align}
        &P(t=0)=P_0\exp(-\frac{z}{H}) \text{,} 
        \label{eq:init_pres}\\
        &\rho(t=0)=\frac{P(t=0)}{gH} \text{, where} \label{eq:init_dens}
\end{align}
$H$ is the scale height of pressure/density and $P_0$, $\rho_0$ ($=P_0/gH$) are the initial values of pressure and density at $z=0$, respectively. The pseudo-entropy $S=P/\rho^\gamma$ has a scale height $H_S(\equiv 1/[\mathrm{d} \ln S/\mathrm{d}z]) = H/(\gamma-1)$. Since $\gamma=5/3$, $H_S>0$ and the equilibrium is convectively stable. 
The degree of stratification is denoted by the Froude number $\mathrm{Fr}$ on the integral scale $\ell_{\mathrm{int}}$ and is given by
\begin{align}
    &\mathrm{Fr}=\frac{v}{N\ell_{\mathrm{int}}}\text{, where} \label{eq:Froude} \\
    &\ell_\mathrm{int}=2\pi\frac{\int{k^{-1}E(k)\mathrm{d}k}}{\int{E(k)}\mathrm{d}k},
\end{align}
\end{subequations}
and $N=\sqrt{g/(\gamma H_S)}$ is the Brunt-V\"{a}is\"{a}l\"{a} oscillation frequency, and $v$ is the rms velocity. The quantity $E(k)$ denotes the velocity power spectrum.

\subsubsection{Turbulent forcing}\label{subsubsec:Turb_forcing}
To force turbulence, we use a spectral forcing method using the stochastic Ornstein-Uhlenbeck (OU) process to model $\bm{a}$ \citep{eswaran1988examination,schmidt2006numerical,Federrath2010A&A}\footnote{The turbulence driving module is publicly available on GitHub \citep{FederrathEtAl2022}.}. The auto-correlation time of the driving is set to roughly match an eddy turnover time on the driving scale. We drive turbulence only on large scales, corresponding to $1\leq k|L/2\pi\leq3$, where $k$ is the magnitude of the wave vector $\bm{k}$. The power is a parabolic function of $k$, peaking at $4\pi/L$, which corresponds to  $\ell_{\mathrm{driv}}=L/2$. We consider two types of forcing in this study--(1) natural mixture and (2) compressive modes only. For a more detailed description of the turbulence driving, we refer the reader to section~2.2.1 of \cite{Mohapatra2022MNRASc}.

\subsubsection{Cooling function}\label{subsubsec:cool_func}
We use the temperature-dependent cooling function  from \cite{Sutherland1993} corresponding to $Z_{\odot}/3$ (a third solar) metallicity. To control the code evolution time step set by $t_{\mathrm{cool}}$, we introduce cutoffs on the cooling rate based on the gas pressure ($P_{\mathrm{cutoff}}$) and temperature ($T_{\mathrm{cutoff}}$). We switch off the gas cooling when the gas pressure or temperature drop below these cutoff values.
We also set a ceiling on the gas density ($\rho_{\mathrm{ceiling}}$) above which we switch off the cooling
The complete cooling function is given by
\begin{subequations}
\begin{equation}
    \mathcal{L}=n_en_i\Lambda(T)\mathcal{H}(T-T_{\mathrm{cutoff}})\mathcal{H}(P-P_{\mathrm{cutoff}})\mathcal{H}(\rho_{\mathrm{ceiling}}-\rho)w(z), \label{eq:cooling_modified}
\end{equation}
where $\mathcal{H}$ is the Heaviside function. We have set $T_{\mathrm{cutoff}}=10^4~\mathrm{K}$, which is also the lower limit of the cooling function in \cite{Sutherland1993}. We fix $P_{\mathrm{cutoff}}=P_0/1000$ and the $\rho_{\mathrm{ceiling}}=500\times\rho_0$.
For faster time steps, we modify the criterion for setting the global time-step of the code $\mathrm{dt}_{\mathrm{code}}$, such that $\mathrm{dt}_{\mathrm{code}}=\mathrm{min}(0.5\times \mathrm{sub}_{\mathrm{factor}} \times t_{\mathrm{cool,min}},\mathrm{dt}_{\mathrm{CFL}})$, where $t_{\mathrm{cool,min}}$ is the minimum cooling time over the domain, $\mathrm{dt}_{\mathrm{CFL}}$ is the code time step set by the Courant-Friedrichs-Lewy criterion and $\mathrm{sub}_{\mathrm{factor}}$ is the subcycling factor which we set to $25$. We refer the reader to appendix C of \cite{Mohapatra2022MNRASc} for a discussion of this implementation. Note that we resolve cooling at most times when we update the internal energy using subcycling.

\subsubsection{Thermal heating rate and shell-by-shell energy balance}\label{subsubsec:shell_balance}
To prevent a runaway cooling flow in the simulation, we implement a shell-by shell balance (in constant $z$ shells) between the net energy lost due to cooling and the net energy added by turbulence and thermal energy input. 
We inject thermal energy into each shell at a rate $Q(z)$ proportional to the local gas density in each shell ($Q \propto \rho$ in Eq. \ref{eq:energy} and $\alpha=1$ in Eq. \ref{eq:t_ti}). However, if the turbulent energy input exceeds the total energy lost in a shell due to cooling, we set $Q(z)=0$ and do not apply any additional cooling. We implement this energy balance at each time step. Mathematically, the heating rate is given by:

\begin{equation}
    Q(z)=\max\left(0,\frac{\rho(x,y,z,t)\int\left(\mathcal{L}- \rho\bm{a}\cdot\bm{v}\right)\mathrm{d}x\mathrm{d}y}{\int\rho\mathrm{d}x\mathrm{d}y}\right)\times w(z).\label{eq:thermal_heating_rate_Q}
\end{equation}

We define the turbulent heating fraction $f_{\mathrm{turb}}$ as
\begin{equation}    f_{\mathrm{turb}}=\frac{\int\rho\bm{a}\cdot\bm{v}\mathrm{d}V}{\int\mathcal{L}\mathrm{d}V}, \label{eq:f_turb}
\end{equation}
\end{subequations}
where we carry out the volume integration over the region defined by $|x|$, $|y|$, $|z|<L/2$.

\subsection{Initial conditions}\label{subsec:init_conditions}
We set our initial conditions to model the dense central regions of CC clusters. We initialise the gas with a constant initial temperature throughout the domain, set to $T_0=1.07\times10^7~\mathrm{K}$, such that the initial sound speed $c_{s0}=500~\mathrm{km/s}$. We set the gas number density $n_0=0.1\mathrm{cm}^{-3}$, so $\rho(t=0)=n_0\mu m_p\exp(-z/H)$ (except for four low-density simulations, where $n_0$ is $2$ times smaller). We drive turbulence on $20~\mathrm{kpc}$ scales, which roughly mimics the size of X-ray cavities seen in the ICM \citep[see e.g.,][for cavity sizes in the MACS clusters sample]{Hlavacek-Larrondo2012MNRAS}. Once turbulence reaches a steady state, the rms velocity of the gas is approximately $250~\mathrm{km/s}$ for our fiducial runs, consistent with the observations by Hitomi in the core regions of the Perseus cluster \citep{hitomi2016}.

The cooling function $\Lambda(T)\propto T^{1/2}$ for free-free cooling at $T\sim 10^7\mathrm{K}$. Since $Q\propto\rho$, this gives $t_{\mathrm{ti}}\approx(10/3)t_{\mathrm{\mathrm{cool}}}$, using $\gamma=5/3$ in \cref{eq:t_ti}.

\subsection{List of simulations}\label{subsec:list_of_models}
We have conducted a total of 16~simulations in this study, which are listed in \cref{tab:sim_params}. By default, our simulations have $512^2\times768$ resolution elements, with $768$ cells along the $z$ axis. Since $L_z=1.5L$, the individual resolution elements (or cells) are all cubical, organised in a uniformly-spaced Cartesian grid. Since we only use the central cubical region with $|x|$, $|y|$, $|z|<L/2$ for the post-processing of our results, the effective resolution is $512^3$.

By default, we drive the natural mixture of turbulent modes \citep[i.e., we do not remove either solenoidal or compressive components of $\bm{a}$; see][]{Federrath2010A&A}. 
Our fiducial set consists of two simulations with different strengths of gravity/stratification (and different $t_{\mathrm{ff}}$) labelled $H1.0$ and $H4.0$ (so the value of $g$ is in the ratio $4:1$). The number following $H$ in the label denotes the scale height of pressure/density in the simulation in code units (i.e., with respect to $L$). We repeat this fiducial set as we vary other simulation parameters in our set. To check the effect of the nature of turbulence forcing, we keep all other parameters fixed but set $\curl{\bm{a}}=\bm{0}$ \citep[compressive forcing; see][]{Federrath2010A&A}. These two runs are indicated by $\zeta0.0$ in the label, where $\zeta$ denotes the fraction of solenoidal modes. In order to vary $t_{\mathrm{mix}}$ while keeping $t_{\mathrm{ti}}$ and $t_{\mathrm{ff}}$ constant, we have two sets of simulations with weak driving and strong driving, denoted as `wdriv' and `sdriv' in the labels, respectively.
Similarly, to check the effect of a longer $t_{\mathrm{ti}}$, we repeat the fiducial set and compressive forcing set of simulations with half the initial density ($n_0=0.05\mathrm{cm}^{-3}$) and pressure, so that the initial temperature still stays the same. This doubles the initial $t_{\mathrm{ti}}$ and $t_{\mathrm{cool}}$. These four runs are marked by `ldens' (low density) in the label. To compare our results directly with previous studies without constant turbulent forcing, we switch off the turbulent forcing and repeat the fiducial set with seed density perturbations at $t=0$. These are marked by `NoTurb' in the run label. 
Finally, to check the convergence of our results, we have two higher resolution versions of our fiducial simulations with $1024^2\times1536$ resolution elements. These simulations are denoted by `HR' in the label.

\begin{table*}
	\centering
	\def\arraystretch{1.2}
	\caption{Simulation parameters and volume-averaged quantities for different runs}
	\label{tab:sim_params}
	\resizebox{\textwidth}{!}{
		\begin{tabular}{lccccccccc} 
			\hline
			Label & Driving & $\mathrm{Fr}$  & $t_{\mathrm{mp}}\ (\mathrm{Gyr})$ & $\mathcal{M}$ & $\mathcal{M}_{\mathrm{comp}}$ & $v\ (\mathrm{km/s})$  &   $t_{\mathrm{ti}}/t_{\mathrm{ff}}$ & $t_{\mathrm{ti}}/t_{\mathrm{mix}}$ & $\sigma_{s,\mathrm{hot}}^2$  \\
			(1) & (2) & (3) & (4) & (5) & (6) & (7) & (8) & (9) & (10)\\
			\hline
			$H1.0$ & Natural & $2.2\pm0.2$ & $1.22$ & $0.64\pm0.01$ & $0.13\pm0.03$ & $255\pm4$ & $3.87\pm0.05$ & $5.92\pm0.08$ & $0.029\pm0.002$ \\
            $H4.0$ & Natural & $5.0\pm0.5$ & NA & $0.64\pm0.01$ & $0.12\pm0.04$ & $259\pm4$ & $2.19\pm0.02$ & $6.92\pm0.04$ & $0.023\pm0.002$ \\
            $\zeta0.0H1.0$ & Compressive & $0.6\pm0.1$ & $0.24$ & $0.40\pm0.01$ & $0.27\pm0.09$ & $172\pm7$ & $6.5\pm0.2$ & $6.2\pm0.5$ & $0.155\pm0.001$ \\
            $\zeta0.0H4.0$ & Compressive & $1.6\pm0.1$ & $0.28$ & $0.40\pm0.01$ & $0.3\pm0.1$ & $167\pm6$ & $2.9\pm0.1$ & $5.4\pm0.4$ & $0.136\pm0.005$ \\
            \hline

            $H1.0$wdriv & Natural & $0.20\pm0.01$ & NA & $0.076\pm0.001$ & $0.015\pm0.004$ & $39\pm1$ & $6.2\pm0.1$ & $1.43\pm0.05$ & $0.006\pm0.001$ \\
            $H4.0$wdriv & Natural & $0.31\pm0.02$ & $1.32$ & $0.047\pm0.004$ & $0.005\pm0.001$ & $24\pm3$ & $2.77\pm0.06$ & $0.8\pm0.1$ & $0.012\pm0.005$ \\
            $H1.0$sdriv & Natural & $2.1\pm0.2$ & $0.19$ & $0.94\pm0.02$ & $0.24\pm0.05$ & $385\pm1$ & $4.0\pm0.1$ & $9.3\pm0.2$ & $0.10\pm0.02$ \\
            $H4.0$sdriv & Natural & $11.0\pm1.0$ & NA & $0.59\pm0.03$ & $0.11\pm0.04$ & $410\pm20$ & $7.1\pm0.1$ & $35.0\pm2.0$ & $0.017\pm0.001$ \\
            
			\hline
            $H1.0$ldens & Natural & $2.5\pm0.1$ & NA & $0.72\pm0.04$ & $0.14\pm0.03$ & $270\pm10$ & $6.6\pm0.2$ & $10.4\pm0.1$ & $0.038\pm0.004$ \\
            $H4.0$ldens & Natural & $5.5\pm0.9$ & NA & $0.65\pm0.02$ & $0.12\pm0.03$ & $266\pm8$ & $4.52\pm0.06$ & $14.6\pm0.2$ & $0.022\pm0.002$ \\
            $\zeta0.0H1.0$ldens & Compressive & $1.2\pm0.09$ & $0.47$ & $0.49\pm0.05$ & $0.29\pm0.08$ & $210\pm20$ & $13.5\pm0.6$ & $16\pm3$ & $0.18\pm0.06$ \\
            $\zeta0.0H4.0$ldens & Compressive & $3.0\pm0.4$ & $0.47$ & $0.49\pm0.03$ & $0.29\pm0.08$ & $210\pm20$ & $6.4\pm0.3$ & $16\pm2$ & $0.17\pm0.04$ \\
            
            \hline
            $H1.0$NoTurb & NA & $0.33\pm0.03$ & NA & $0.10\pm0.01$ & $0.019\pm0.004$ & $52\pm1$ & $6.63\pm0.03$ & $2.07\pm0.06$ & $0.007\pm0.001$ \\
            $H4.0$NoTurb & NA & $0.36\pm0.05$ & $0.52$ & $0.05\pm0.01$ & $0.013\pm0.009$ & $23\pm4$ & $2.80\pm0.04$ & $0.8\pm0.2$ & $0.013\pm0.003$ \\
            
            \hline
            $H1.0$HR & Natural & $2.2\pm0.1$ & $1.45$ & $0.70\pm0.02$ & $0.15\pm0.05$ & $266\pm6$ & $3.6\pm0.1$ & $5.9\pm0.1$ & $0.043\pm0.003$ \\
            $H4.0$HR & Natural & $4.8\pm0.1$ & NA & $0.66\pm0.02$ & $0.12\pm0.03$ & $261\pm7$ & $2.19\pm0.03$ & $7.03\pm0.08$ & $0.025\pm0.003$ \\
            \hline
	\end{tabular}}
	\justifying \\ \begin{footnotesize} Notes: Column 1 shows the simulation label. The number following $H$ denotes the scale height of the initial pressure/density profile in code-units. We show the type of turbulence driving in column 2. In column 3, we show the average Froude number $\mathrm{Fr}$ of the simulations. The fourth column shows the time at which multiphase gas condenses out of the hot phase through thermal instability for a simulation. We denote it as `NA' if there is no multiphase gas condensation in the particular simulation. In columns 5 and 6, we show the volume-weighted rms Mach number and its compressive component $\mathcal{M}_\mathrm{comp}$, respectively. In column 7, we show the volume-weighted standard deviations of velocity $v$. We show the average value of the ratio between the thermal instability timescale $t_{\mathrm{ti}}$ and important dynamical time scales - the free-fall time scale $t_{\mathrm{ff}}$ and the turbulent mixing time scale $t_{\mathrm{mix}}$ in columns 8 and 9, respectively. Finally, in column 10, we show $\sigma^2_{s,\mathrm{hot}}$, the square of the standard deviations of the logarithms of density of the hot phase. All time-averaged statistics in columns 3, 5, 6, 7, 8, 9 and 10 are averaged for $t\leq t_{\mathrm{mp}}$ for runs in which multiphase gas forms. Movies of simulations are available at this \href{https://youtube.com/playlist?list=PLuaNgQ1v_KMZlkKXdB7hcaQ7-hb0hmY7G}{playlist}. \end{footnotesize} 
	
\end{table*}

\section{Results and discussion}\label{sec:results-discussion}
In this section, we present and discuss the results of our simulations. We have run all our simulations till $t_{\mathrm{end}}=2.344~\mathrm{Gyr}$. Thermal instability leads to cold gas condensing out of the hot phase in $8$ out of our $14$~simulations. For runs that form multiphase gas, we define the time at which cold ($T\lesssim2\times10^4~\mathrm{K}$) gas first forms (when the cold gas mass fraction $m_{\mathrm{cold}}/m_{\mathrm{tot}}>0.01\%$) as $t_{\mathrm{mp}}$ and list it in column 4 of \cref{tab:sim_params}. We have also listed some time and volume-averaged statistics in \cref{tab:sim_params}, such as $\mathrm{Fr}$, the rms Mach number $\mathcal{M}$, the rms velocity $v$, the average value of the ratio between important time-scales $t_{\mathrm{ti}}/t_{\mathrm{ff}}$ and $t_{\mathrm{ti}}/t_{\mathrm{mix}}$, and the square of logarithmic-density ($s$) fluctuations $\sigma_{s,\mathrm{hot}}^2$ in columns 3, 5, 6, 7, 8 and 9, respectively. For runs that do not form multiphase gas, these quantities are averaged over the last $120~\mathrm{Myr}$ of the simulation. For runs that form multiphase gas, these averages are calculated in the $120~\mathrm{Myr}$ just before $t_{\mathrm{mp}}$, but after the first $100~\mathrm{Myr}$, so that there is some time for turbulence to grow\footnote{Note that we expect turbulence to grow and reach a steady state in roughly $2$--$3$ eddy turnover time-scales \citep{Federrath2010A&A}, which corresponds to $150$--$250~\mathrm{Myr}$ for our fiducial set of runs. For some of our runs, this time-scale is longer than $t_{\mathrm{mp}}$. For such runs, we calculate the time and volume-averaged quantities in the last $25~\mathrm{Myr}$ just before $t_{\mathrm{mp}}$, to reduce the effect of unsaturated turbulence-evolution on the time-averaging.}.


We begin this section by briefly discussing some key statistical properties of the gas in the fiducial set and the compressive forcing set of runs. These are crucial to understanding the second part of our study, where we vary the simulation parameters such as the strength of the turbulence forcing and the cooling rate. In the later subsections we move our focus to the non-linear evolution of thermal instability in the system and how it is affected by the different parameter choices.

\subsection{Fiducial and compressive forcing  runs}\label{subsec:fid_case}
\subsubsection{Projection maps perpendicular to the stratification}
\begin{figure*}
		\centering
	\includegraphics[width=2.0\columnwidth]{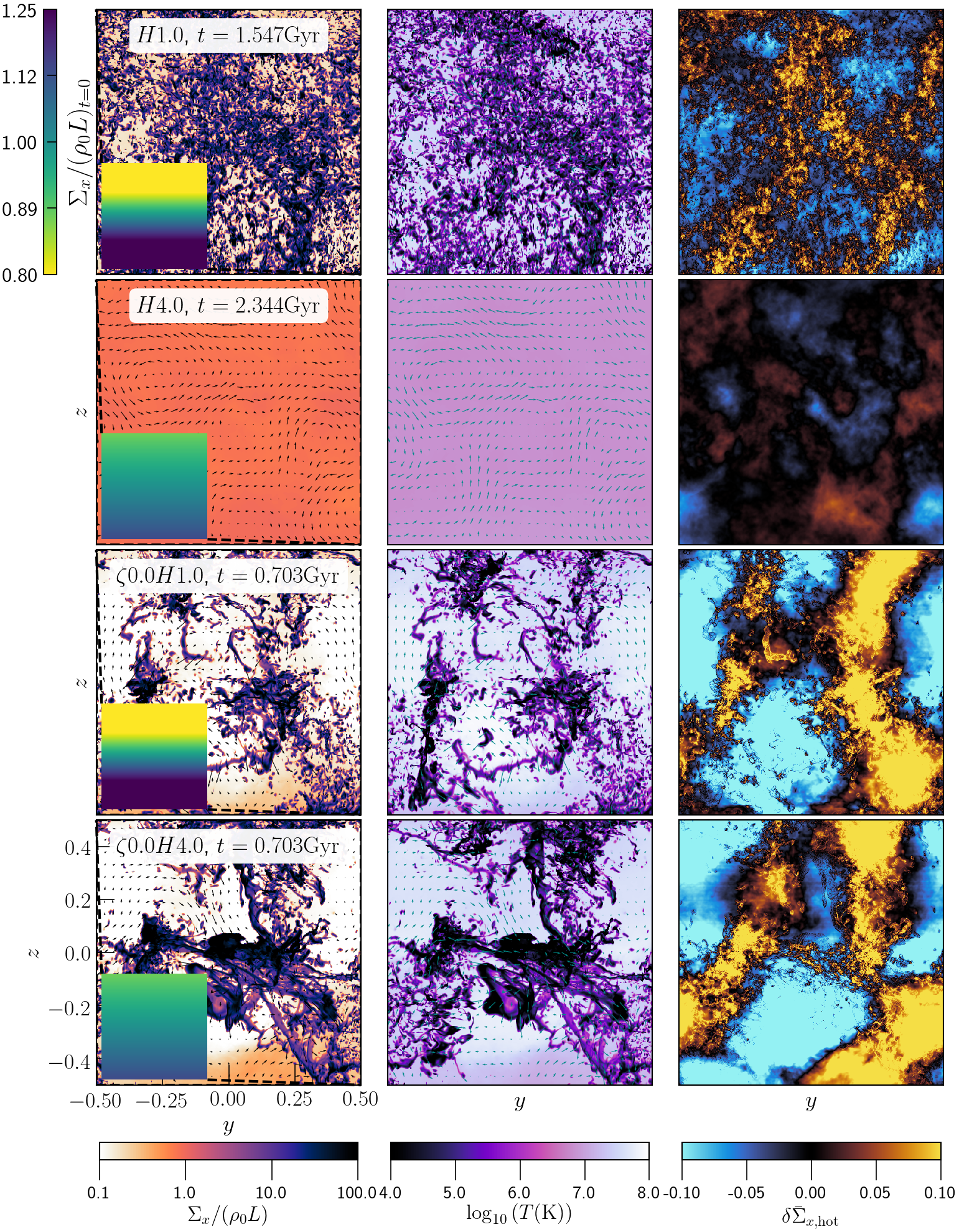}	
	\caption[Projections of density, temperature hot phase density fluctuations for fiducial runs]{Density (volume-weighted), temperature (mass-weighted) and normalised column density fluctuations in the hot phase ($T>10^6~\mathrm{K}$), integrated along the $x$-axis for our fiducial and compressive driving sets of runs. The insets in column 1 show the column density of the gas at $t=0$. Cold gas forms through condensation from the hot phase for all runs except the $H4.0$ run. This produces large variations in the gas density and temperature. The compressive forcing runs produce large-scale cold filamentary clouds.}
	\label{fig:projplots-fid}
\end{figure*}
Three of these runs form cold gas through thermal instability, but the $H4.0$ run doesn't. In \cref{fig:projplots-fid}, we show the projections of gas density (volume-weighted, first column), temperature (mass-weighted, second column) and column density fluctuations (after dividing out the $xy$-averaged density profile) in the hot phase ($T\geq10^6~\mathrm{K}$, third column). These snapshots are plotted when the runs have the maximum mass fraction of cold gas ($m_{\mathrm{cold}}/m_{\mathrm{tot}}$) and at $t=t_{\mathrm{end}}$ for the $H4.0$ run. The insets in column~1 show the projections of gas density at $t=0$. Clearly, the runs with $H=1.0$ have stronger gradients in the initial density than the runs with $H=4.0$.

Thermal instability produces large variations in density, with much stronger variations compared to the initial density gradient. In all runs that form multiphase gas, the dense regions correspond to cooler gas and the rarer regions correspond to hotter gas, as expected. For the $H1.0$ run, the cold clouds are misty, i.e., they are small in size and occur throughout the simulation domain. In comparison, the compressive driving runs show many large clouds, with size $\sim\ell_\mathrm{driv}=20~\mathrm{kpc}$. These results are similar to what we observed for different forcing runs in simulations without gravity in \cite{Mohapatra2022MNRASc}.

For the $H4.0$ run, the net variations in density and temperature are much smaller compared to the other runs. Column density fluctuations in the hot phase are also much weaker for this run. For the other runs, we find that the regions with cold gas (in column 2) are associated with strong, positive fluctuations in the column density in the hot phase (in column 3). Such features are also observed in multi-wavelength observations of the ICM (see e.g.,\citealt{Werner2013ApJ,Anderson2018A&A,Baek2022ApJ}). In our simulations, the spatial overlap between the different phases could be either due to turbulent mixing with the cooler gas making the hot phase denser or the cold gas could have directly formed from these dense regions of the hot gas, which have shorter cooling time (since $t_{\mathrm{cool}}\propto\rho^{-1}$).

\subsubsection{Time-evolution of volume-averaged quantities}\label{subsubsec:time-evol-diff-driving}
\begin{figure}
		\centering
	\includegraphics[width=\columnwidth]{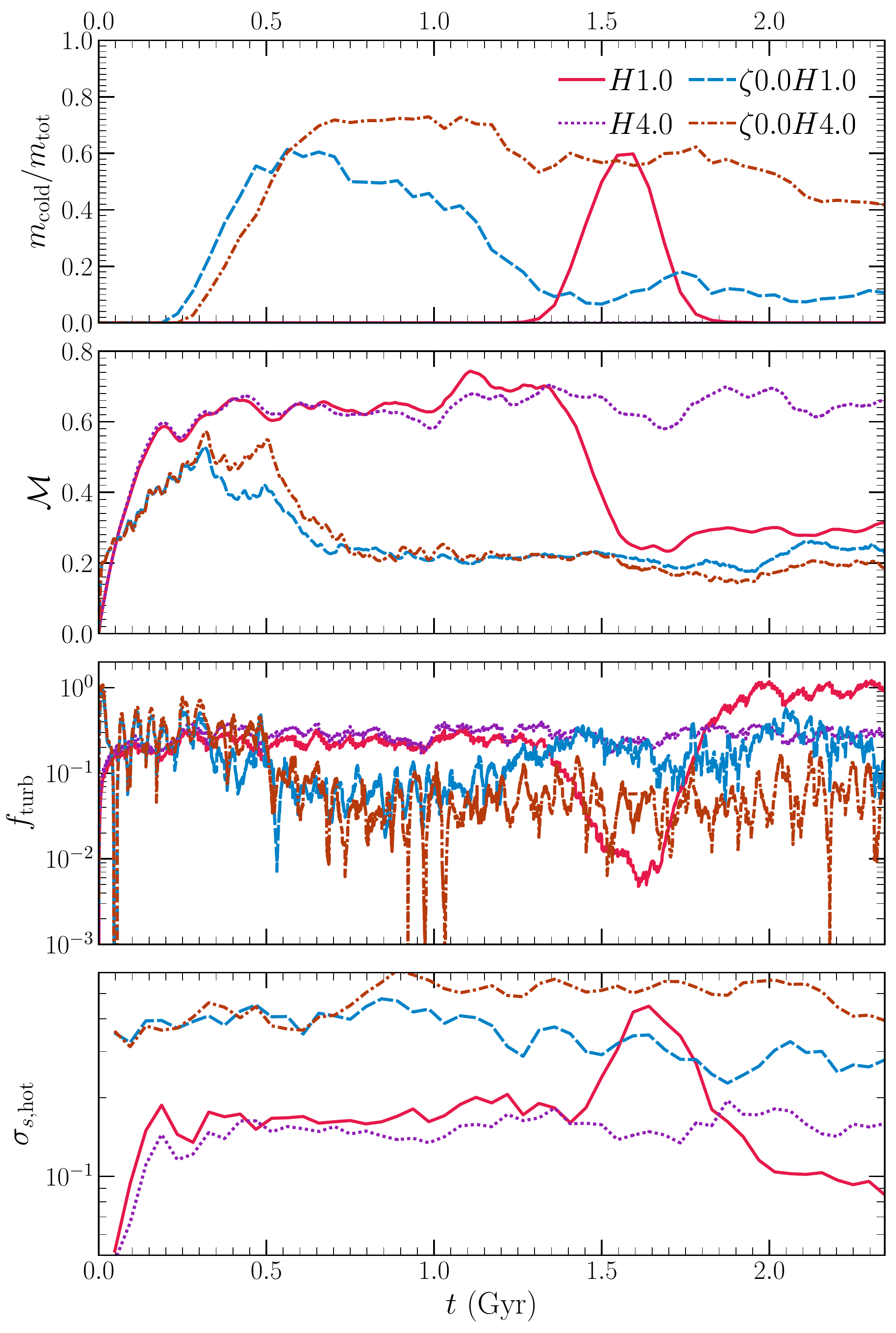}	
	\caption[Time evolution of relevant quantities for fiducial runs]{Time evolution of the cold-gas mass fraction (first row), volume-weighted rms Mach number (second row), turbulent heating fraction $f_{\mathrm{turb}}$ (third row), and amplitude of logarithmic density fluctuations in the hot phase ($T>10^6~\mathrm{K}$, fourth row), for our fiducial and compressive driving sets of runs.}
	\label{fig:time-evolution-fid}
\end{figure}
In \cref{fig:time-evolution-fid}, we show the time-evolution of the mass fraction of cold gas ($T\lesssim2\times10^4~\mathrm{K}$) in the first row, the volume-averaged $\mathcal{M}$ in the second row, $f_{\mathrm{turb}}$ (defined in \cref{eq:f_turb}) in the third row and the standard deviation of logarithmic density of the hot-phase $\sigma_{s,\mathrm{hot}}$ in the fourth row. 

Cold gas forms at different times ($t_{\mathrm{mp}}$) for the three different runs. The time $t_{\mathrm{mp}}$ is clearly affected by the driving, multiphase gas condensation occurs much earlier for the compressive forcing runs. This is due to the stronger seed density fluctuations generated by the compressive forcing, as seen in the fourth row of \cref{fig:time-evolution-fid}. The ratio $m_{\mathrm{cold}}/m_{\mathrm{tot}}$ initially increases, reaches a maximum value and then decreases with time. The rate of decrease in $m_{\mathrm{cold}}/m_{\mathrm{tot}}$ is much faster for the runs with stronger gravity (i.e., $H=1.0$), since the cold clumps being heavier than the ambient hot gas, fall faster to the negative $z$ boundary. 

At initial times, $\mathcal{M}$ for all runs reaches values of $0.5$--$0.7$. The turbulent heating fraction $f_{\mathrm{turb}}$ is approximately a few $\times\;10\%$. However, for the runs forming multiphase gas, we find that both $\mathcal{M}$ and $f_{\mathrm{turb}}$ decrease at $t=t_{\mathrm{mp}}$. By design, the turbulent forcing amplitude remains the same throughout the duration of the simulation. Cold-gas condensation is associated with the production of fast-cooling dense gas at intermediate temperatures ($2\times10^4~\mathrm{K}\lesssim T\lesssim10^6~\mathrm{K}$), which increases the cooling rate. This is compensated by an increase in the heating rate since we impose energy balance in $z$-shells. The rarer hot-phase gas is heated more (because $\mathcal{L}\propto\rho^2$, $Q\propto\rho$), which increases $c_\mathrm{s}$ and decreases $\mathcal{M}$.  

At late times, the simulation reaches a steady state at a lower $\mathcal{M}$ but higher $f_{\mathrm{turb}}$. The atmosphere is hotter and has a smaller net cooling rate, such that $f_{\mathrm{turb}}$ increases. For the $H1.0$ run, after the removal of extra mass, the turbulent heating alone is sufficient to balance the reduced steady-state cooling rate ($f_{\mathrm{turb}}=1$). 

Among the two fiducial runs ($H1.0$ and $H4.0$), the hot-gas density fluctuations are slightly larger for the $H1.0$ run for $t<t_{\mathrm{mp}}$. This happens because the $H1.0$ run is more strongly stratified ($\mathrm{Fr}$ listed in column~3 of \cref{tab:sim_params}) compared to the $H4.0$ run. \cite{Mohapatra2020,Mohapatra2021MNRAS} showed that for weak and moderate levels of stratification ($\mathrm{Fr}\gtrsim1$) the density fluctuations increase with increasing stratification (decreasing $\mathrm{Fr}$) for fixed $\mathcal{M}$ and driving. These larger seeds lead to multiphase condensation developing in the $H1.0$ run (and a slightly shorter cooling time, whose effect we discuss later), whereas they do not develop in the $H4.0$ run.

The hot-gas density fluctuations show a sharp increase at $t\gtrsim t_{\mathrm{mp}}$ for the $H1.0$ run---bringing its value closer to the amplitudes for the compressive forcing runs. Clearly, the density fluctuations due to multiphase condensation are much larger than those due to stratified turbulence at $t<t_{\mathrm{mp}}$. Using unstratified multiphase turbulence simulations in \citet[][figure~6 and section~3.5]{Mohapatra2022MNRASc}, we showed that these larger fluctuations are due to the strong compressive velocities during cold-gas condensation and the baroclinicity of a multiphase turbulent system. 

\subsubsection{Mach number, temperature and density distributions}\label{subsubsec:mach_temp_dens_PDFs}
\begin{figure*}
		\centering
	\includegraphics[width=2\columnwidth]{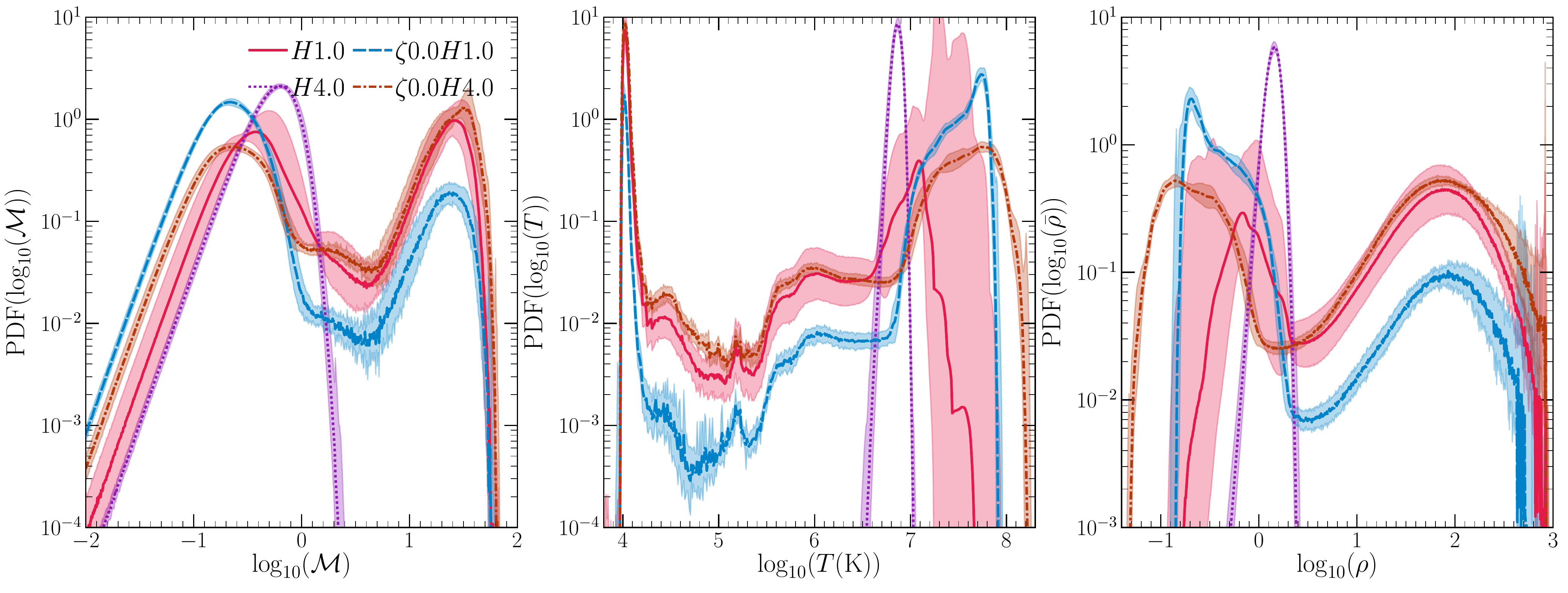}	
	\caption[PDFs of density, temperature and Mach number for fiducial runs]{The mass-weighted probability distribution functions of Mach number (left), temperature (middle), and density (right), for our fiducial and compressive driving sets of runs. The $H4.0$ run does not form cold gas and shows a single peak in all distributions, while the other three runs that form cold gas show two strong peaks, corresponding to the hot and cold phases. The hot-phase gas is hotter (by about an order of magnitude) for the two compressive forcing runs.}
	\label{fig:1D-pdfs-fid}
\end{figure*}

In \cref{fig:1D-pdfs-fid} we show the mass-weighted probability distribution functions (PDFs) of the Mach number (first column), temperature (second column) and gas density (third column) for our fiducial and compressive driving sets of runs. The PDFs for the three multiphase runs are averaged from $1.4~\mathrm{Gyr}$ to $1.64~\mathrm{Gyr}$ and for the single-phase $H4.0$ run, they are averaged from $1.4~\mathrm{Gyr}$ till $t_{\mathrm{end}}$. We show the $1-\sigma$ spread in PDF values as shaded regions. 
The runs forming multiphase gas show two strong peaks in all three PDFs, whereas the $H4.0$ run shows a single peak. The two peaks correspond to the hot and cold phases. 

The hot phase is subsonic ($\mathcal{M}_{\mathrm{hot}}<1$) for all four runs, as is expected from ICM observations (\citealt{hitomi2016}, see \citealt{Simionescu2019SSRv} for a review). The high $\mathcal{M}$ peak corresponds to the supersonic cold-phase gas, which has much smaller sound speed. Since we use the same forcing scheme to drive turbulence in all four runs, the shapes of the distributions of $\mathcal{M}$ are quite similar for $\mathcal{M}\lesssim1$. The small offsets can be explained by differences in the temperature/sound speed among the different runs. 

In the temperature PDFs, we observe a strong cold-phase peak at $T_{\mathrm{cutoff}}=10^4~{\mathrm{K}}$ and the hot-phase peak at $T\sim10^7$--$10^8~\mathrm{K}$. The features in the PDF between these two peaks correspond to the shape of the cooling curve that we use. The temperature of the hot-phase peak is higher for the compressive forcing runs.

In the density PDFs, the low-density peak corresponds to the hot phase and the high-density peak to the cold phase. The hot-phase gas has much lower density for the compressive forcing runs, while the density of the cold-phase peak is similar. Thus, the ratio between the densities of the phases $\chi=\rho_{\mathrm{cold}}/\rho_{\mathrm{hot}}$ is much larger for compressive forcing. This is caused by strong converging and diverging motions on the driving scale \citep{Schmidt2009A&A,Federrath2010A&A,Seta2022MNRAS}. 
For the $H4.0$ run, the density PDF is log-normal with a power-law tail at low densities. The low-density tail is a known feature of the PDFs when the adiabatic index $\gamma>1$, also reported in \cite{Passot1998PhRvE,Federrath2015MNRAS,Mohapatra2020}.

\subsubsection{Density-temperature phase diagram} \label{subsubsec:dens_temp_phase_diagram}
\begin{figure*}
		\centering
	\includegraphics[width=2\columnwidth]{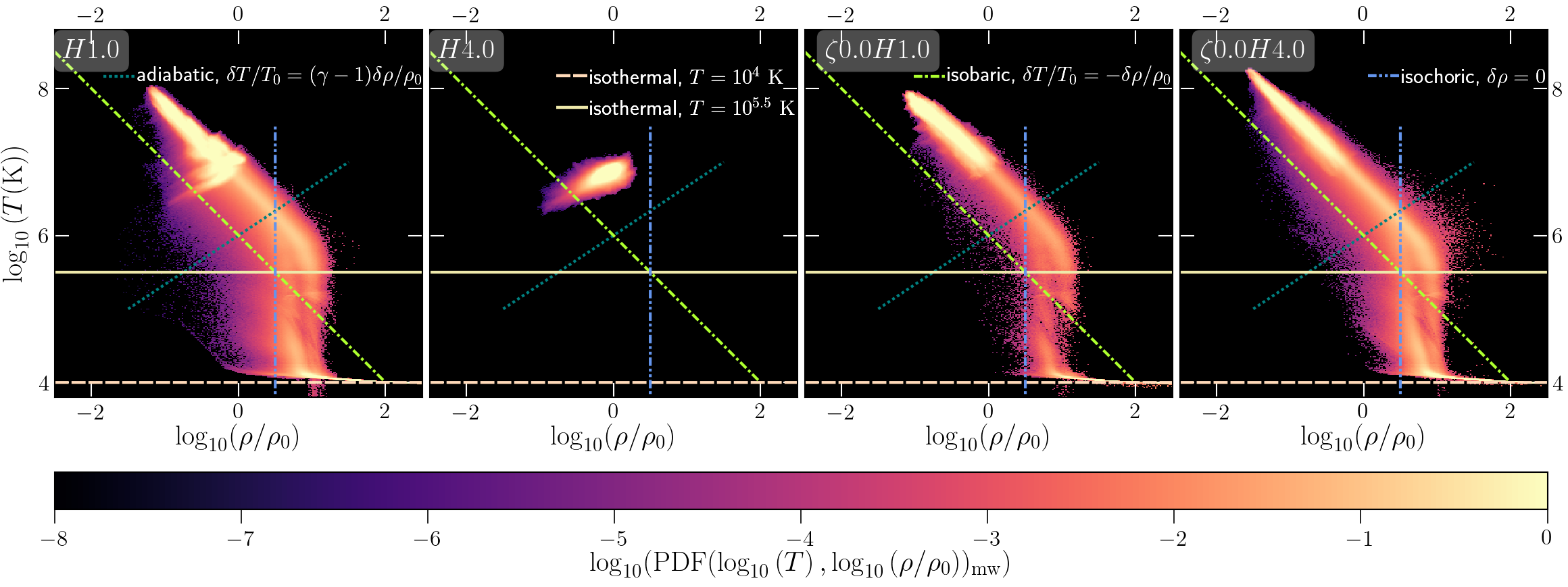}	
	\caption[Joint PDF of density and temperature for fiducial runs]{The mass-weighted 2D PDFs of $T$ vs $\rho$ for our fiducial and compressive driving sets of runs. The single-phase $H4.0$ run shows a mixture of isobaric and adiabatic modes. The three multiphase runs show an isobaric hot phase ($T>10^6~\mathrm{K}$), an isochoric intermediate phase  ($2\times10^4~\mathrm{K}<T<10^6~\mathrm{K}$) and an isothermal cold phase ($T\lesssim2\times10^4~\mathrm{K}$).}
	\label{fig:dens-temp-pdf-fid}
\end{figure*}

In \cref{fig:dens-temp-pdf-fid}, we show the joint mass-weighted PDFs of the logarithms of temperature and density, temporally averaged over the same duration as the 1D PDFs in \cref{fig:1D-pdfs-fid}. The different lines show the nature of fluctuations: adiabatic ($\delta T/T_0\propto(\gamma-1)\delta\rho/\rho_0$), isothermal ($\delta T=0$) at $10^{5.5}~\mathrm{K}$ and $T_{\mathrm{cutoff}}=10^4~\mathrm{K}$, isobaric ($\delta T/T_0=-\delta\rho/\rho_0$) and isochoric ($\delta\rho/\rho_0$). From a theoretical viewpoint, understanding the nature of fluctuations is important to calculate the growth rate of thermal instability through the different fluctuation modes \citep{Das2021MNRAS}. They are also useful to compare with observations. For instance, \cite{zhuravleva2018} inferred the mode of perturbations from X-ray observations of the ICM. 

In our single-phase $H4.0$ run, the fluctuations are composed of isobaric and adiabatic components. This is in agreement with the stratified turbulence simulations (without radiative cooling) of \cite{Mohapatra2020}, where we showed that unstratified turbulence produces adiabatic fluctuations, and the fraction of isobaric fluctuations increases with increasing strength of the stratification.

For the multiphase runs, we observe some clear trends in the PDFs --- the hot phase ($10^6$--$10^8~\mathrm{K}$) is isobaric, the intermediate temperatures are isochoric, with a drop in temperature around $10^{5.5}$--$10^6~\mathrm{K}$ and the cold phase is approximately isothermal at $T_{\mathrm{cutoff}}$. We reported the same features in the temperature-density joint PDFs in \citet[][figure~5]{Mohapatra2022MNRASc}, so they are not strongly affected by the stratification. 

The isochoric drop at $T~\sim10^{5.5}$--$10^6~\mathrm{K}$ is associated with the peak of $\Lambda(T)$, where $t_{\mathrm{cool}}<t_{\mathrm{cs}}$. The cooling time for the gas at intermediate temperatures is quite short and such gas may not be able to attain pressure equilibrium. However, some of this pressure drop could be due to our lack of resolution of the cooling length ($\ell_{\mathrm{cool}}=\min(c_\mathrm{s} t_{\mathrm{cool}})$). Recent high-resolution simulations of multiphase systems such as \cite{Fielding2020ApJ,Abruzzo2022arXiv221015679A} argue that this could be due to lower spatial resolution in large-scale boxes, which do not resolve $\ell_{\mathrm{cool}}$. While resolving $\ell_{\mathrm{cool}}$ is important to model the properties of the cold phase after it forms, it is not necessary to determine when or where it forms. In this study we mainly focus on the latter part, so we do not expect our results to strongly depend on resolution. We have checked our results for convergence in \cref{app:convergence_test}. The TNG50 simulations \citep{Nelson2020MNRAS,Ramesh2023MNRAS}, which track the cold gas better than our fixed-grid simulations, do not show this isochoric drop. However, this could be partly due to the orders of magnitude variation in halo pressure in TNG50 halos (therefore the sharp isochoric temperature drop is not as clear) whereas the vertical extent of our simulation box is much smaller to have a large pressure variation.

\subsubsection{Evolution of the $z$-profile of entropy}\label{subsubsec:entropy_z_profile_evol}

\begin{figure*}
		\centering
	\includegraphics[width=2\columnwidth]{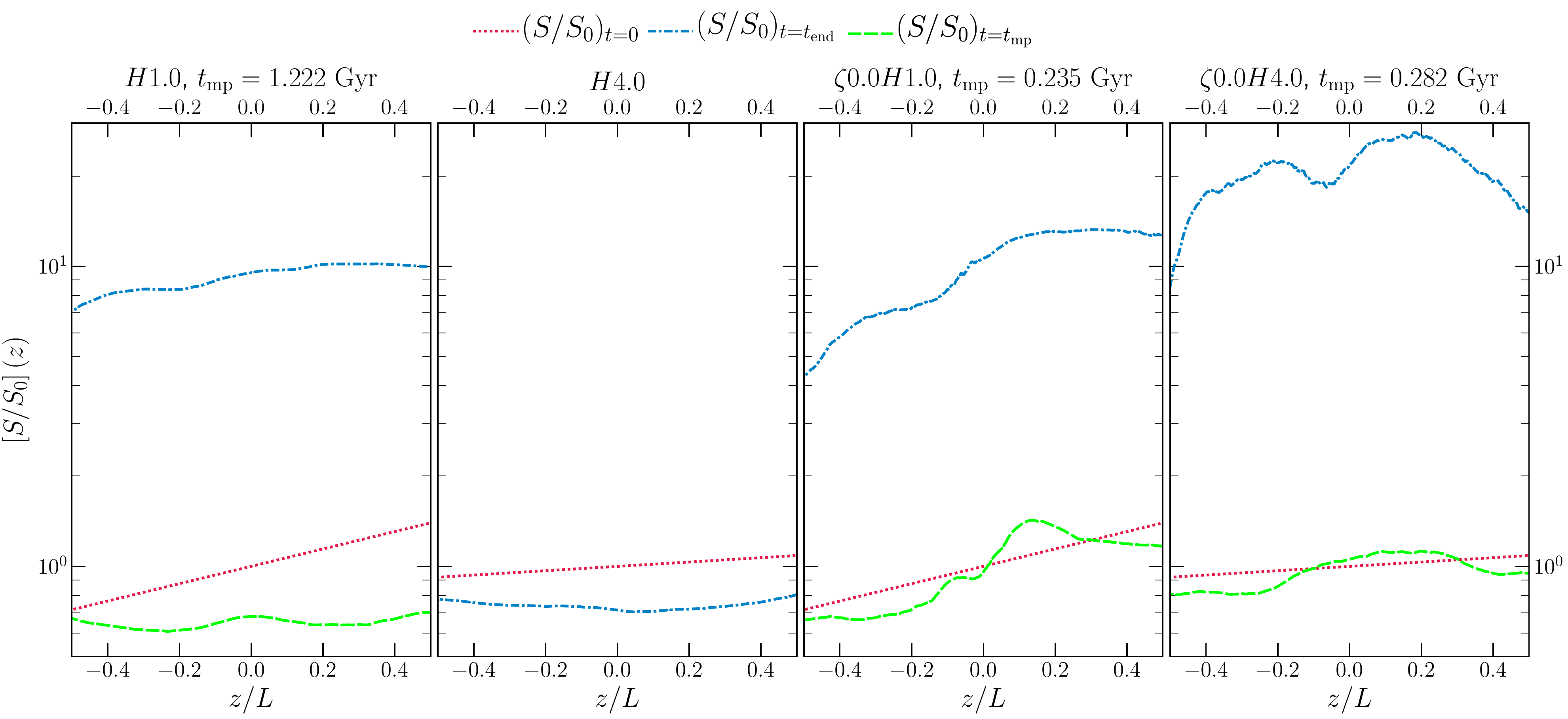}	
	\caption[Entropy profiles for fiducial runs]{The vertical profiles of entropy of the hot phase ($T\gtrsim10^6~\mathrm{K}$, averaged along the $xy$ plane) at $t=0$ (red dotted line) and $t=t_{\mathrm{end}}$ (blue dash-dotted line) for our fiducial and compressive driving sets of runs. We also show the entropy profile at $t=t_{\mathrm{mp}}$ (green dashed line, when cold gas has just started forming) for runs that form multiphase gas.}
	\label{fig:entr-profz-fid}
\end{figure*}

Theoretical studies such as \cite{Voit2017ApJ} report that the large-scale entropy gradient is important to thermal instability. They propose that halos in thermal balance (applicable to our setup) with a shallower entropy gradient are more susceptible to condensation. In \cref{fig:entr-profz-fid}, we show the $z$-shell averaged entropy profiles ($\left[S/S_0\right](z)$, where $S_0=P_0/\rho_0^\gamma$) of the hot gas ($T\gtrsim10^6~\mathrm{K}$) for our fiducial and compressive forcing sets of runs at $t=0$ and $t=t_{\mathrm{end}}$. For the three runs that form multiphase gas, we also plot the entropy profile at the onset of multiphase condensation ($t_{\mathrm{mp}}$, denoted in the titles of the respective columns). 

For the $H1.0$ run, the entropy gradient is steep at $t=0$, but it flattens out around the onset of multiphase condensation ($t=t_{\mathrm{mp}}$). This is due to turbulent mixing, which mixes the low- and high-entropy regions together and makes the entropy gradient disappear. After cold gas condenses and moves out of the box through the bottom $z$ boundary, at $t=t_{\mathrm{end}}$ the entropy increases by almost an order of magnitude. We find that the gas has redeveloped a weak entropy gradient at this time.

The single-phase $H4.0$ run starts out with a much weaker entropy gradient compared to the $H1.0$ run. Despite starting out with a flatter entropy gradient, this run never forms multiphase gas. By $t=t_{\mathrm{end}}$, its entropy gradient also disappears and its entropy value is slightly larger than that for the $H1.0$ run just before condensation. 

The two compressive forcing runs form multiphase gas fairly quickly. Our snapshots just before thermal condensation show that the initial entropy profiles have large-scale variations even within the first $\sim300~\mathrm{Myr}$ of the simulations. By this time, the turbulence is still developing, such that a large-scale entropy gradient has not been lost to the mixing. By $t=t_{\mathrm{end}}$, the average entropy for both runs increases by an order of magnitude. Unlike the $H1.0$ run, we still observe a strong entropy gradient for the $\zeta0.0H1.0$ run. The large-scale entropy profile shows a very disturbed state for the $\zeta0.0H4.0$ run due to strong large-scale perturbations induced by the compressive forcing, which are not moved out of the box by the weaker gravity.

In summary, we find that a smaller initial entropy gradient (larger $H$) does not necessarily imply better thermal stability of the halo. The entropy profile can be strongly modified by large-scale turbulence, which can remove the initial gradients, given enough time ($H1.0$ and $H4.0$ runs). Further, the different amplitudes of density fluctuations also play a key role---larger fluctuations can seed multiphase condensation even when the entropy gradient is steep.

\subsubsection{Evolution of $z$-profiles of important timescales}\label{subsec:timescales_zprof_evol}
\begin{figure*}
		\centering
	\includegraphics[width=2\columnwidth]{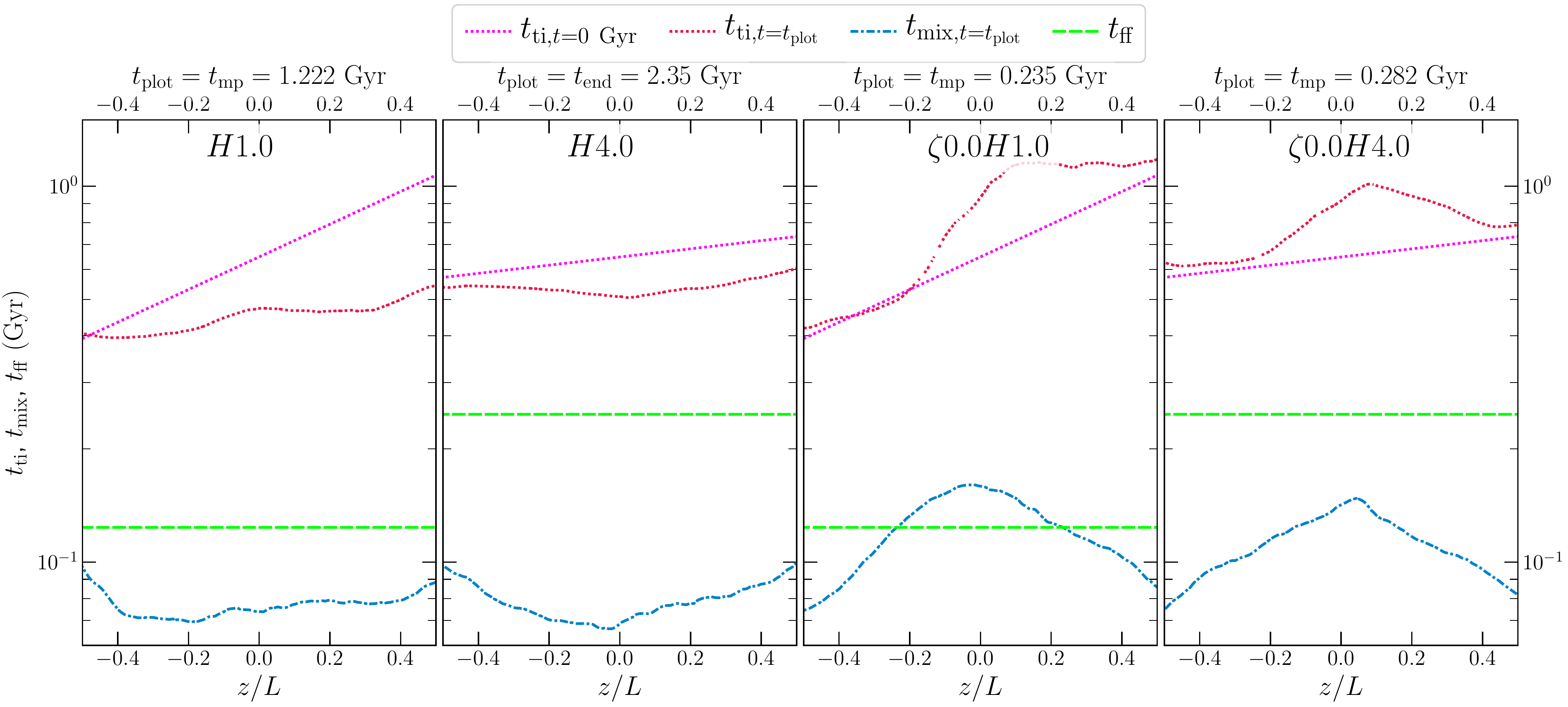}	
	\caption[Important timescales for fiducial runs]{The variation of important timescales for the hot-phase gas ($T\gtrsim10^6~\mathrm{K}$)--$t_{\mathrm{ti}}$, $t_{\mathrm{ff}}$ and $t_{\mathrm{mix}}$ averaged in shells parallel to the $z$ axis for our fiducial and compressive driving sets of runs at $t=0$. For runs in which multiphase gas forms through thermal instability, we also show $t_{\mathrm{ti}}$ and $t_{\mathrm{mix}}$ at the onset of multiphase condensation (at $t_{\mathrm{plot}}=t_{\mathrm{mp}}$, denoted in the column titles). For the single phase runs, we show these timescale profiles at $t_\mathrm{plot}=t_\mathrm{end}$.}
	\label{fig:timescales-fid}
\end{figure*}

Following the discussion on the role played by the entropy profile, we now move our attention to the $z$ shell-averaged values of the three important timescales of the system $t_{\mathrm{ti}}$, $t_{\mathrm{mix}}$ and $t_{\mathrm{ff}}$ (defined in \cref{subsec:imp_timescales}). The ratio between these timescales is expected to play a key role in the thermal stability of the system and has been studied in both theoretical \citep[e.g.,][]{sharma2012thermal,mccourt2012,Gaspari2018ApJ}, numerical \citep[e.g.,][]{prasad2015,Beckmann2019A&A,Butsky2020ApJ} and observational \citep[e.g.,][]{Voit2015ApJ,Olivares2019A&A} studies. In \cref{fig:timescales-fid}, we show these quantities for the hot phase ($T\geq10^6~\mathrm{K}$) at $t=0$ and at the onset of multiphase condensation ($t_{\mathrm{plot}}=t_{\mathrm{mp}}$). For the runs that do not form multiphase gas, we set $t_\mathrm{plot}=t_\mathrm{end}=2.344~\mathrm{Gyr}$.

We start with an isothermal profile, so at $t=0$, $t_{\mathrm{ti}}\propto\rho^{-1}$ (see eq.~\ref{eq:t_cool}). It varies exponentially with $z$, with a scale height $H$. The free-fall time $t_{\mathrm{ff}}$ is a constant throughout space and time, since we fix $\bm{g}$ to a constant value. 

For the $H1.0$ run, the $z$-gradient of $t_{\mathrm{ti}}$ flattens and its value decreases slightly, following the same trend as the evolution of the entropy profile shown in \cref{fig:entr-profz-fid}. Around the time when cold gas starts condensing out of the medium ($t=t_{\mathrm{mp}}$), $t_{\mathrm{ti}}/t_{\mathrm{ff}}=3.87\pm0.05$ and $t_{\mathrm{ti}}/t_{\mathrm{mix}}=5.92\pm0.08$. This medium satisfies the instability criterion ($t_{\mathrm{ti}}/t_{\mathrm{ff}}\lesssim10$) proposed by \cite{sharma2012thermal} and produces multiphase gas. However, \cite{Gaspari2018ApJ} argue that when $t_{\mathrm{ti}}/t_{\mathrm{mix}}>1$, turbulent mixing should be able to stop multiphase gas from developing. However, this criterion does not correctly predict the outcome of the $H1.0$ simulation. 
By $t=t_{\mathrm{end}}$, cold gas condenses out and falls through the bottom $z$-boundary. In the new steady state, the hotter and rarer atmosphere has $t_{\mathrm{ti}}\sim10~\mathrm{Gyr}$, $t_{\mathrm{ti}}/t_{\mathrm{ff}}\approx80$ (see movie of timescale profiles evolution in supplementary material or at this \href{https://youtu.be/CoalcA9DCpI}{link}) and is stable against undergoing further thermal condensation.

For the single-phase $H4.0$ run, the evolution of $t_{\mathrm{ti}}$ is similar to that of the $H1.0$ run, but its average value is slightly larger. The ratio $t_{\mathrm{ti}}/t_{\mathrm{ff}}=2.19\pm0.02$ and $t_{\mathrm{ti}}/t_\mathrm{mix}=6.92\pm0.04$. For this run, the criterion by \cite{Gaspari2018ApJ} correctly predicts that multiphase condensation does not occur in this system, while the \cite{sharma2012thermal} prediction does not hold true. 

The amplitude of seed density fluctuations plays a key role in determining whether the systems undergo condensation. The $H4.0$ run has weaker seed density perturbations compared to the $H1.0$ run (see row~4 in \cref{fig:time-evolution-fid}) and a slightly larger $t_{\mathrm{ti}}/t_\mathrm{mix}$. The relatively faster mixing of the weaker seeds successfully prevents cold gas from condensing out. The two compressive forcing runs have much larger seed density perturbations. Despite having $t_{\mathrm{ti}}/t_{\mathrm{mix}}=6.2\pm0.5$ and $5.4\pm0.4$ at $t=t_{\mathrm{mp}}$ for the $\zeta0.0H1.0$ and $\zeta0.0H4.0$ runs, respectively, they both form multiphase gas. At $t=t_{\mathrm{end}}$, the $\zeta0.0H1.0$ run has a similar value of $t_{\mathrm{ti}}$ as the $H1.0$ run, albeit with larger variations due to the compressive forcing. In comparison, the $\zeta0.0H4.0$ run reaches a larger $t_{\mathrm{ti}}$ in steady state, but a similar $t_{\mathrm{ti}}/t_{\mathrm{ff}}\approx100$.

\subsection{Effect of weaker/stronger forcing}\label{subsec:diff_forcing}
\begin{figure}
		\centering
	\includegraphics[width=\columnwidth]{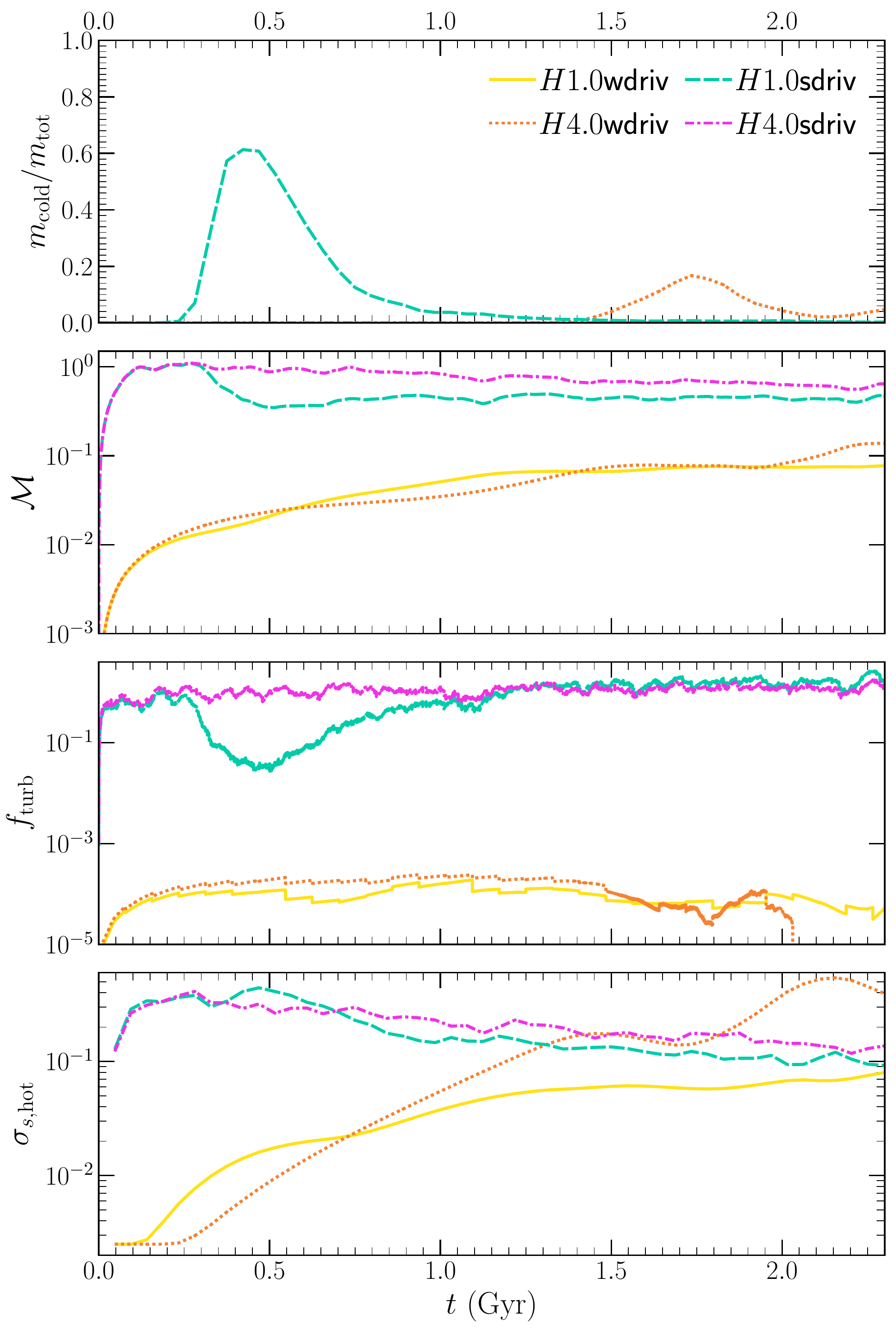}	
	\caption[Time evolution of relevant quantities for different Mach number runs]{Similar to \cref{fig:time-evolution-fid}, but for our weak- and strong-driving set of runs. We observe contrasting trends in the development of multiphase condensation with increasing stratification for the weak and strong driving sets of runs.}
	\label{fig:time-evolution-diffmach}
\end{figure}

\begin{figure*}
		\centering
	\includegraphics[width=2\columnwidth]{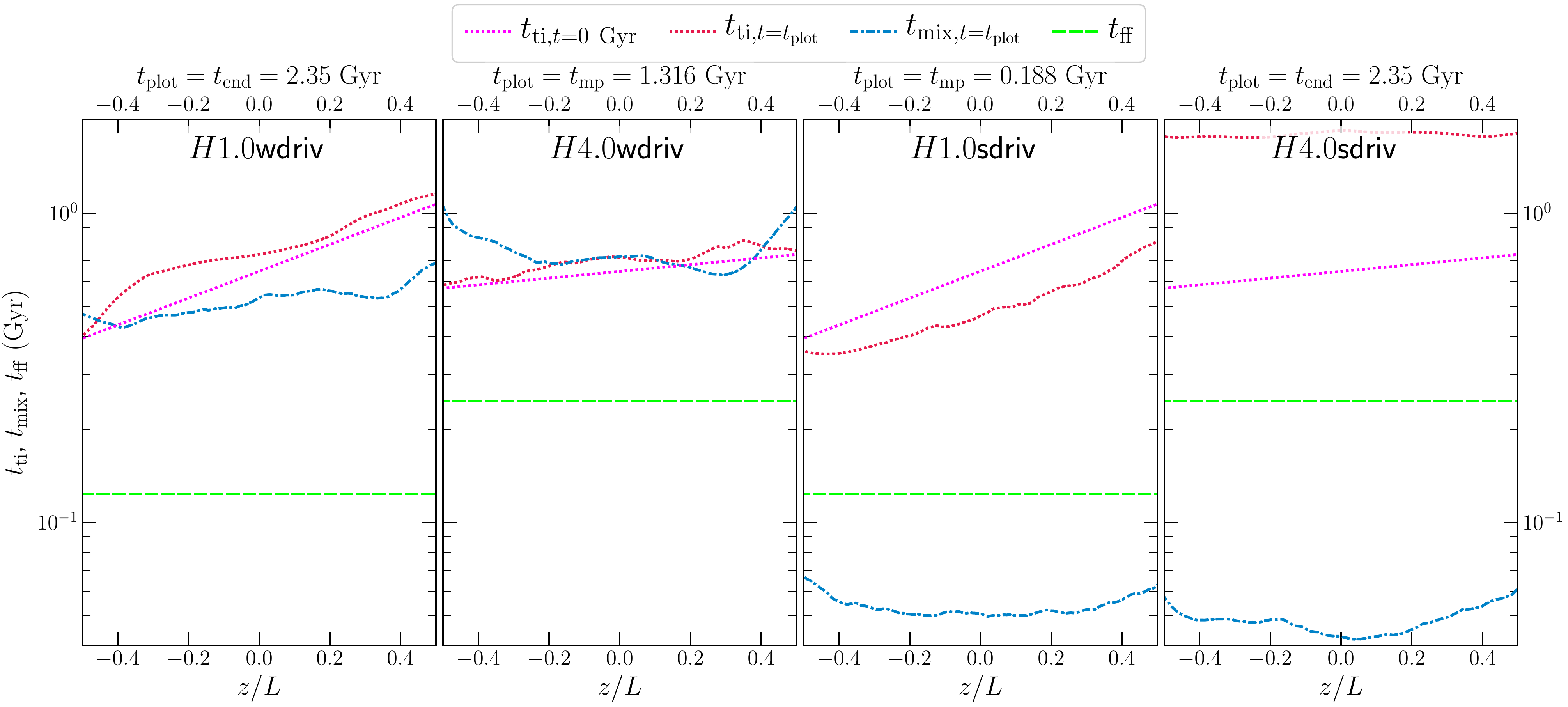}	
	\caption[Important timescales for different Mach number runs]{Similar to \cref{fig:timescales-fid}, but for our for our weak (wdriv) and strong (sdriv) driving sets of runs. For weak driving, the weaker stratification run forms multiphase gas, while for strong driving, the stronger stratification run shows multiphase gas.}
	\label{fig:timescales-diffmach}
\end{figure*}

Considering the importance of the turbulence driving for the formation of multiphase gas seen in the previous subsections, here we analyse four more runs, where we vary the strength of the turbulence forcing. In steady state, $v\sim20$--$40~\mathrm{km/s}$ for the two `wdriv' runs and $\sim400~\mathrm{km/s}$ for the two `sdriv' runs. Similar to \cref{fig:time-evolution-fid}, in \cref{fig:time-evolution-diffmach}, we show the time evolution of the $m_{\mathrm{cold}}/m_{\mathrm{tot}}$, $\mathcal{M}$, $f_{\mathrm{turb}}$ and $\sigma_{\mathrm{s,hot}}$. We present the $z$ shell-averaged profiles of important time-scales (for the hot phase) in \cref{fig:timescales-diffmach}.

Out of the four runs, $H4.0$wdriv and $H1.0$sdriv form multiphase gas, whereas $H1.0$wdriv and $H4.0$sdriv do not. 
First we focus our discussion here on the `wdriv' set of runs. Due to the weak forcing, these two runs are the most comparable to thermal instability studies that do not explicitly drive turbulence \citep[such as][]{sharma2012thermal,Choudhury2019}.\footnote{For a direct comparison with \cite{sharma2012thermal,Choudhury2019}, we have also conducted two simulations `$H1.0$NoTurb' and `$H4.0$NoTurb' where we only introduce seed density fluctuations and do not drive turbulence explicitly. The results from these simulations are consistent with the corresponding `wdriv' set of runs and are also in agreement with the aforementioned studies of thermal instability.}

The turbulent eddy turnover time for these two runs is around $0.5$--$0.7~\mathrm{Gyr}$. Due to the weaker forcing, turbulence is strongly stratified, with $\mathrm{Fr}\ll1$. In this regime, \citet[][figure~5]{Mohapatra2021MNRAS} showed that density fluctuations decrease with increasing stratification, due to strong buoyancy forces limiting motions in the $z$-direction. 

This is clearly observed in our simulations (fourth row of \cref{fig:time-evolution-diffmach}) as the density fluctuations are smaller for the $H1.0$wdriv run compared to those for the $H4.0$wdriv run (for $t\gtrsim0.8~\mathrm{Gyr}$). The weaker seed fluctuations are thus unable to induce multiphase condensation in the $H1.0$wdriv run, even though $t_{\mathrm{ti}}/t_\mathrm{ff}=6.2\pm0.1$. In \cref{fig:timescales-diffmach}, we find that the weak forcing is unable to significantly modify the initial profile of $t_{\mathrm{ti}}$ by $t=t_{\mathrm{end}}$, unlike the fiducial set, which flattened the $z$-profiles of $t_{\mathrm{ti}}$ (and entropy). 

For the $H4.0$wdriv run, $t_{\mathrm{mix}}\sim t_\mathrm{ti}$ around $1.316~\mathrm{Gyr}$, when the driven turbulence is expected to reach a steady state.
Due to the weak turbulent mixing between the $z$-shells, most of the cold gas condensation occurs from the lower half of the box, which has a smaller initial $t_{\mathrm{ti}}$ (see movies of simulation in supplementary material or at this \href{https://youtube.com/playlist?list=PLuaNgQ1v_KMbQpTMc6_nu5deeclJ-LN2B}{playlist link}). 
Compared to the $\zeta0.0H4.0$ run, $t_{\mathrm{ti}}\sim2$--$5~\mathrm{Gyr}$ at $t=t_{\mathrm{end}}$, which is an order of magnitude smaller. Thus, for weaker driving, the system does not lose as much mass to condensation during the simulation period of $2.344~\mathrm{Gyr}$. 

The trend in the two `sdriv' runs are similar to what we observe for the fiducial set---out of the two, the more strongly stratified $H1.0$sdriv run forms multiphase gas, while the weakly-stratified $H4.0$sdriv run doesn't. There are a few differences---the initial density fluctuations are larger for the $H1.0$sdriv run, so the multiphase gas forms much earlier compared to the $H1.0$ run from the fiducial set even before the $z$-profile of $t_{\mathrm{ti}}$ is flattened by turbulent mixing. 

Before the onset of multiphase condensation, the amplitude of fluctuations in the $H1.0$sdriv and $H4.0$sdriv runs around $t=0.2~\mathrm{Gyr}$ are similar \citep[in agreement with expectations from][for $\mathcal{M}\sim1$]{Mohapatra2021MNRAS}. The key difference between the two is the shorter average $t_\mathrm{ti}$ in $H1.0$sdriv. Although $t_\mathrm{ti}/t_\mathrm{mix}=9.3\pm0.2$, it is still unable to stop multiphase gas from developing. In the $H4.0$sdriv run, the turbulent heating due to the strong driving ($v=410\pm20~\mathrm{km/s}$) is more than sufficient to offset the cooling ($f_\mathrm{turb}\gtrsim1$). The gas heats up with time, showing a gradual decrease in $\mathcal{M}$ and a larger value of $t_{\mathrm{ti}}$ at $t=t_{\mathrm{end}}$.


\subsection{Effect of weaker cooling}\label{subsec:weak_cooling}
\begin{figure}
		\centering
	\includegraphics[width=\columnwidth]{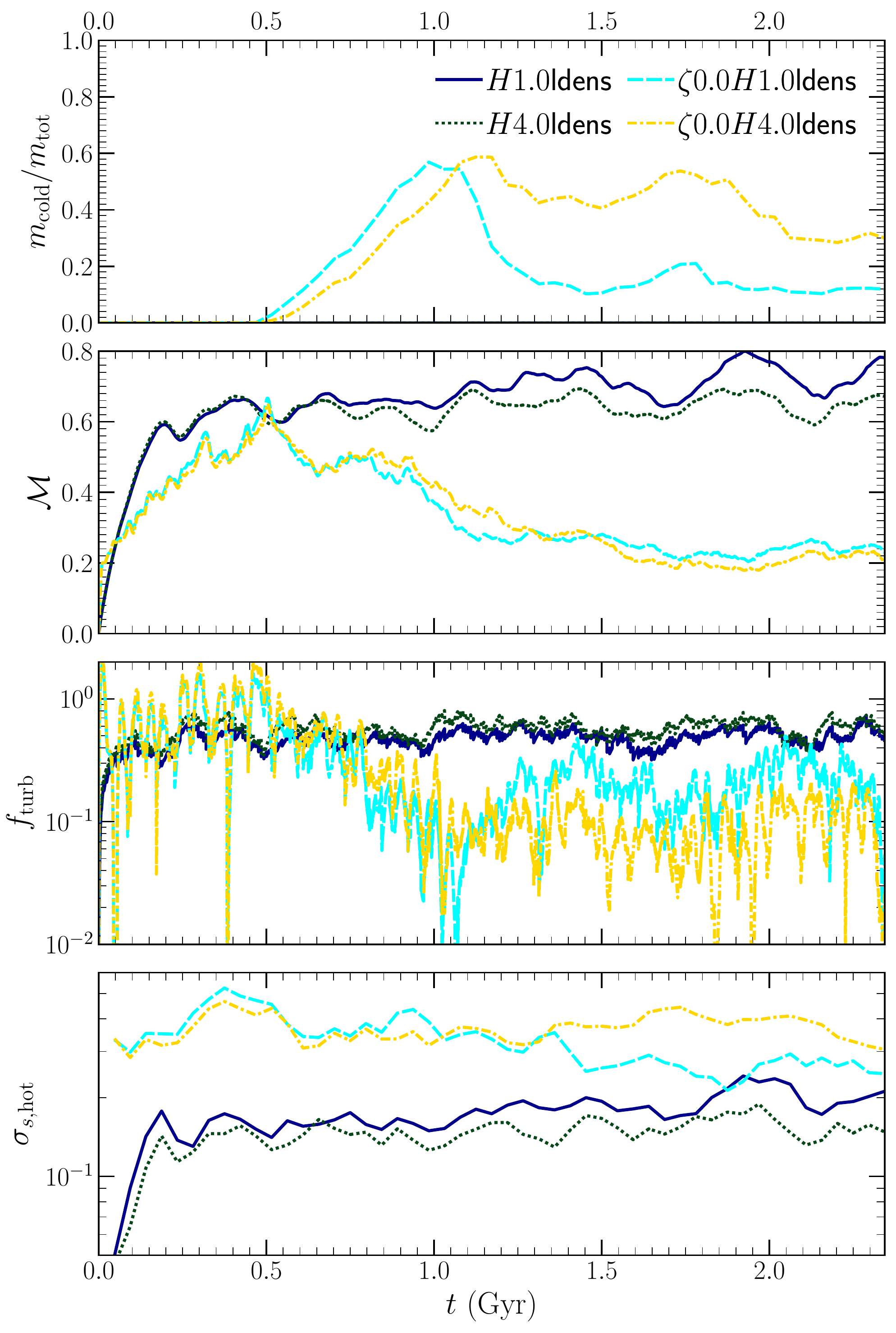}	
	\caption[Time evolution of relevant quantities for different Mach number runs]{Similar to \cref{fig:time-evolution-fid}, but for our lower initial density (weaker cooling) set of runs. Only the compressive forcing runs form multiphase gas, albeit at a much later time compared to their fiducial set counterparts.}
	\label{fig:time-evolution-lowdens}
\end{figure}
\begin{figure*}
		\centering
	\includegraphics[width=2\columnwidth]{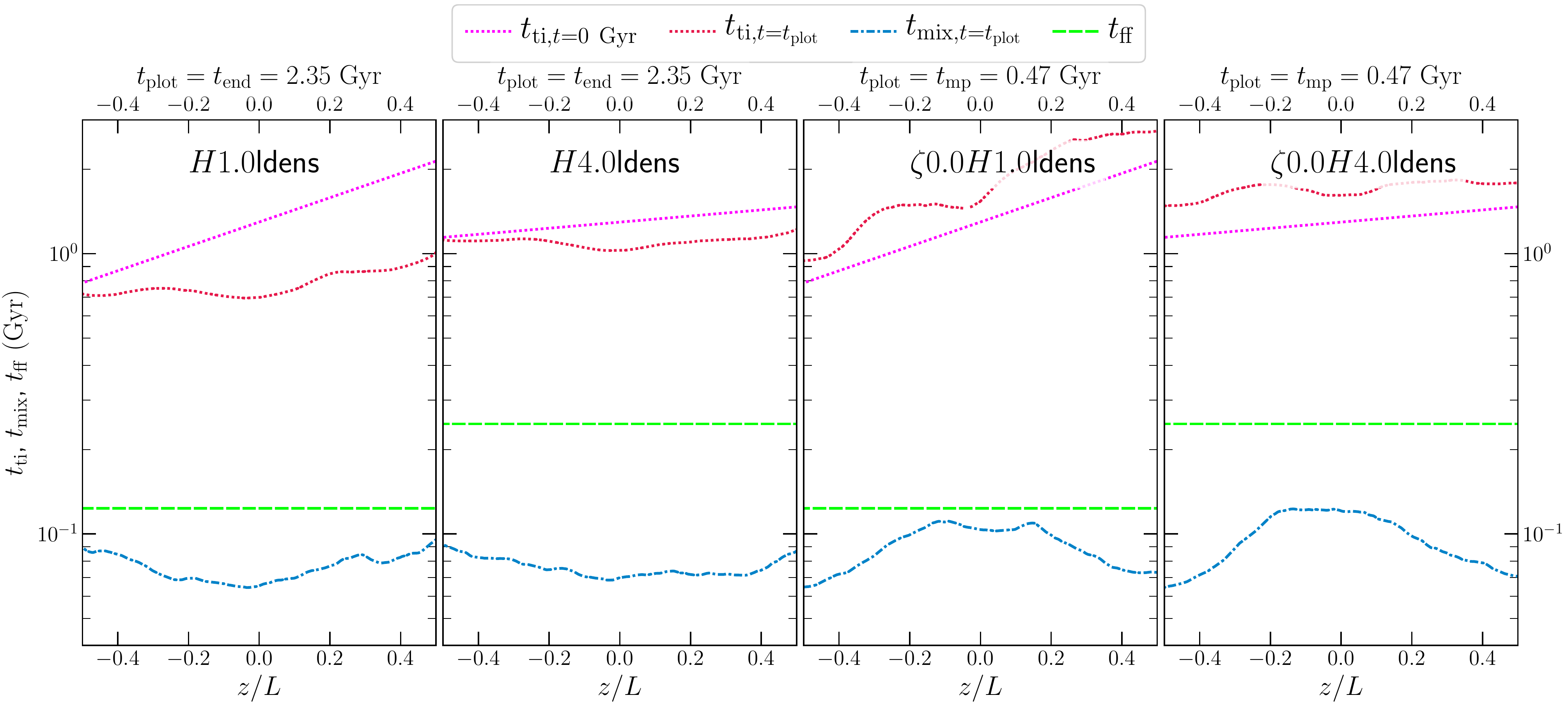}	
	\caption[Important timescales for lower density runs]{Similar to \cref{fig:timescales-fid}, but for our `lowdens' set of runs. The initial density is half compared to the fiducial set, which doubles $t_\mathrm{ti}$. Only the compressive forcing runs form multiphase gas.}
	\label{fig:timescales-lowdens}
\end{figure*}

For the runs described in this subsection, we lower $\rho_0$ and $P_0$ by half compared to the fiducial set (so initial $T$ is fixed). This doubles $t_{\mathrm{ti}}$, while $t_{\mathrm{ff}}$ and $t_{\mathrm{mix}}$ are unaffected. We show the time evolution of relevant quantities in \cref{fig:time-evolution-lowdens} and the $z$ shell-averaged timescale profiles in \cref{fig:timescales-lowdens}. These are low-density (or longer $t_{\mathrm{ti}}$) counterparts to figures \ref{fig:time-evolution-fid} and \ref{fig:timescales-fid} for the fiducial set. 

We find that only the two compressive forcing runs form multiphase gas, while the natural forcing runs do not. Since $t_\mathrm{cool}$ and $t_\mathrm{ti}$ are doubled, $t_{\mathrm{mp}}\sim500~\mathrm{Myr}$ is also doubled for these runs compared to $\sim250$--$300~\mathrm{Myr}$ for the fiducial compressive set with the same parameters. These two runs show a clear decrease in $\mathcal{M}$ around $t_\mathrm{mp}$ associated with the hot phase becoming hotter. 
Since the cooling is weaker, $f_\mathrm{turb}$ is larger, roughly by a factor of two for all the low-density runs compared to their fiducial counterparts. The fraction $f_\mathrm{turb}\approx30\%$ for the natural forcing runs and $50$--$100\%$ for the compressive forcing runs for $t<t_\mathrm{mp}$. For $t>t_\mathrm{mp}$, $f_\mathrm{turb}$ decreases, similar to what we observe for the fiducial set. 

In \cref{fig:timescales-lowdens}, we find that turbulent mixing flattens the $z$ profiles of $t_\mathrm{ti}$ for both the natural driving runs. The average $t_\mathrm{ti}/t_\mathrm{ff}=6.6\pm0.2$, $t_\mathrm{ti}/t_\mathrm{mix}=10.4\pm0.1$ for $H1.0$ldens and $t_\mathrm{ti}/t_\mathrm{ff}=4.52\pm0.06$, $t_\mathrm{ti}/t_\mathrm{mix}=14.6\pm0.2$ for $H4.0$ldens run. The larger value of these ratios compared to the fiducial set, ensures that multiphase condensation does not occur in either of these runs. 

For the compressive forcing runs, the average  values of $t_\mathrm{ti}/t_\mathrm{ff}=13.5\pm0.6$, $t_\mathrm{ti}/t_\mathrm{mix}=16\pm3$ for $\zeta0.0H1.0$ldens and $t_\mathrm{ti}/t_\mathrm{ff}=6.4\pm0.3$, $t_\mathrm{ti}/t_\mathrm{mix}=16\pm2$ for $\zeta0.0H4.0$ldens. Both of these ratios are much larger than $1$. Both \cite{sharma2012thermal} and \cite{Gaspari2018ApJ} models would predict the $\zeta0.0H1.0$ldens run to not produce multiphase gas, contrary to what we find\footnote{Although the $z$ shell-averaged values of $t_\mathrm{ti}/t_\mathrm{ff}$ and $t_\mathrm{ti}/t_\mathrm{mix}$ are large, these ratios can become much smaller in dense, locally compressed regions produced by the compressive forcing.}. However, the large density fluctuations due to the compressive forcing grow before either mixing or buoyancy can prevent them from becoming multiphase. By $t=t_\mathrm{end}$, $t_\mathrm{ti}\sim10$--$30~\mathrm{Gyr}$ similar to that of their fiducial counterparts, despite their longer initial $t_\mathrm{ti}$. Thus, $\sigma_s$, $t_\mathrm{ti}/t_\mathrm{ff}$ and $t_\mathrm{ti}/t_\mathrm{mix}$ determine the final value of $t_\mathrm{ti}$ rather than the initial value of $t_\mathrm{ti}$.

\section{Summary of the timescale ratios and their implications} \label{sec:timescale_ratio_summary}
\begin{figure}
		\centering
	\includegraphics[width=\columnwidth]{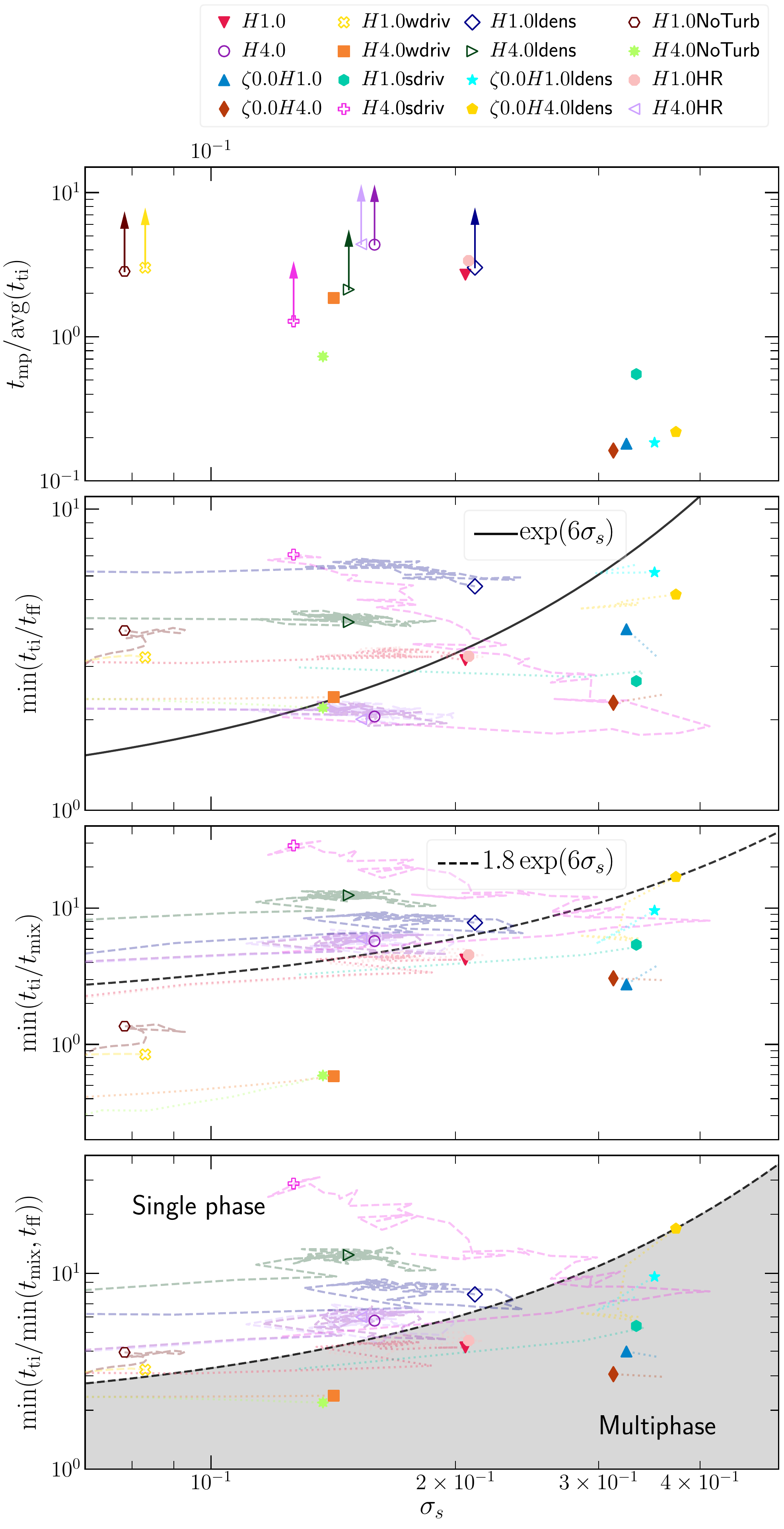}	
	\caption[min(tti/tff) and min(tti/tmix) vs dens fluc]{\emph{First row:} Scatter plot of the time taken to form multiphase gas normalised by the $z$ shell-averaged thermal instability time scale ($t_\mathrm{mp}/t_\mathrm{ti}$) vs.~the standard deviation in the logarithm of gas density ($\sigma_s$) for all our runs. The filled points show runs that form multiphase gas, while the unfilled points show runs that remain single phase till $t=t_\mathrm{end}$. For the latter set of runs, we show the lower limits to the ratio, denoted by the upward facing arrows in the symbols. \emph{Second row:} The minimum value of the ratio of $t_{\mathrm{ti}}$ to the $z$ shell-averaged free-fall time scale ($t_{\mathrm{ff}}$) $t_{\mathrm{ti}}/t_{\mathrm{ff}}$ with the same $x$~axis. \emph{Third row:} Similar to the upper panel, but we show $\min(t_\mathrm{ti}/t_\mathrm{mix})$, the minimum value of the ratio between the $z$ shell-averaged $t_{\mathrm{ti}}$ and the turbulent mixing time scale ($t_{\mathrm{mix}}$) instead along the $y$-axis. \emph{Fourth row:} Here we show $\min(t_\mathrm{ti}/\min(t_\mathrm{mix},t_\mathrm{ff}))$, using the minimum of $t_\mathrm{ff}$ and $t_\mathrm{mix}$ in the denominator instead. The black line corresponds to the condensation curve described in \cref{eq:condensation_curve_tti_by_tff}. For the third and fourth rows, the black dashed line is given by \cref{eq:min_tti_min_tmix_tff}. It clearly separates between the single phase and multiphase runs in the fourth row. The coloured dashed lines show the time evolution of these ratios as a function of $\sigma_s$ till $t=\min(t_\mathrm{mp},t_\mathrm{end})$}.
	\label{fig:min_tti_tff_vs_sigs}
\end{figure}

Here we summarise our results from all our simulations and discuss them in the broader context of the conditions that lead to multiphase condensation in the halo gas. In \cref{fig:min_tti_tff_vs_sigs} we show the time taken to form multiphase gas normalised by the thermal instability time scale ($t_\mathrm{mp}/t_\mathrm{ti}$) (first row), minimum values of the ratios $t_\mathrm{ti}/t_\mathrm{ff}$ (second row), $t_\mathrm{ti}/t_\mathrm{mix}$ (third row) and $t_\mathrm{ti}/\min(t_\mathrm{ff},t_\mathrm{mix})$ (fourth row)\footnote{Note that we calculate the minimum value of these ratios using the $z$-shell averaged values of $t_\mathrm{ti}$ and $t_\mathrm{mix}$ instead of calculating their minimum values over the entire domain. This makes our results directly comparable to the radial profiles of the timescales obtained from observations. The local variations in $t_\mathrm{ti}$ are mostly due to density fluctuations, which are captured well by $\sigma_s$.} as a function of the standard deviation of logarithmic density (normalised) for all of our 16 simulations. For runs that form multiphase gas, we show these values just before $t_\mathrm{mp}$ and plot them as filled data-points. For the runs that do not form multiphase gas, we plot the ratios at $t=t_\mathrm{end}$ using unfilled data-points. The coloured dashed lines show the time evolution of these quantities as a function of $\sigma_s$ prior to multiphase condensation (or the end of the simulation). 

\subsection{Time taken to form multiphase gas}\label{subsec:time_t_mp}
Out of our 16 simulations, 9 form multiphase gas. For the 7 simulations that remain single phase till $t=t_\mathrm{end}$, we plot $t_\mathrm{end}/\mathrm{avg}(t_\mathrm{ti})$ as a lower limit to $t_\mathrm{mp}/\mathrm{avg}(t_\mathrm{ti})$, in the first row of \cref{fig:min_tti_tff_vs_sigs}. The single phase simulations are generally concentrated to the upper left part of the figure, whereas the multiphase simulations are to the bottom right. This denotes that larger density fluctuations aid the formation of multiphase gas. Among the runs that form multiphase gas, we find that we can further divide them into three sub-groups. The forcing in the the four compressive driving runs and the strong driving $H1.0$sdriv generates large density fluctuations ($\sigma_s \gtrsim 0.3$) and the gas forms localised high-density pockets with a short cooling time. The multiphase gas forms in $t_\mathrm{mp}\lesssim0.5 t_\mathrm{ti}$ for these simulations. The remaining four multiphase runs form cold gas at $t_\mathrm{mp}\simeq t_\mathrm{ti}$. We note that the runs with stronger turbulence ($H1.0$ and $H1.0$HR) have stronger density fluctuations but form multiphase gas later compared to the runs with weak or no turbulent forcing ($H4.0$wdriv and $H4.0$NoTurb). This highlights that turbulence driving generates stronger density fluctuations but turbulence mixing slows the onset of multiphase condensation. On the other hand, in the absence of mixing the amplitude of density fluctuations keeps growing with time for the $H4.0$wdriv and $H4.0$NoTurb runs till $t=t_\mathrm{mp}$ (see fourth panel of fig.~\ref{fig:time-evolution-diffmach}).


\subsection{A condensation curve for the formation of multiphase gas}\label{subsec:condensation_curve_multiphase}
In this subsection, we first discuss how the predictions of thermal instability criteria proposed by \cite{sharma2012thermal} and \cite{Gaspari2018ApJ} hold for our set of simulations. We also attempt to construct a modified condensation curve based on these two criteria for our simulations, taking into account the local variation in $t_{\mathrm{ti}}$ due to density fluctuations, as well as the log-normal shape of the density distribution (and consequently $t_\mathrm{cool}$, since $t_\mathrm{cool}\propto\rho^{-1}$) before multiphase condensation occurs (e.g.~see the density PDF for the $H4.0$ run in fig.~\ref{fig:1D-pdfs-fid}). Since condensation is a local phenomenon, i.e., dense pockets of gas with a short ratio of the timescales can condense out even when the atmosphere is globally stable \citep[also seen in][]{Choudhury2019}, we consider the minimum value of these timescales in our criterion. The densest regions would have gas density $\rho_\mathrm{max}\sim \mean{\rho}\exp(c_1\sigma_s)$, where $c_1$ is a positive constant. As $t_\mathrm{cool}\propto\rho^{-1}$, $\min(t_{\mathrm{cool}})\sim \mean{t_{\mathrm{cool}}}\times\exp(\mathrm{-c_1}\sigma_s)$. Similar to \cite{Voit2021ApJ}, we use an exponential condensation curve that depends on $\sigma_s$, and which takes into account these local variations in $t_{\mathrm{ti}}$ (or $t_\mathrm{cool}$) due to density fluctuations. 

\subsubsection{The importance of $t_\mathrm{ti}/t_{\mathrm{ff}}$}\label{subsubsec:t_ti_tff_criterion}
\cite{sharma2012thermal} propose the criterion $t_\mathrm{ti}/t_\mathrm{ff}\lesssim10$ for the onset of multiphase condensation. This is satisfied in all our simulations, barring the $\zeta0.0H1.0$ldens run. Yet 8 out of the 15~simulations do not form multiphase gas, indicating that turbulent mixing has a significant effect on the conditions required for multiphase condensation \citep[also discussed in][]{banerjee2014turbulence,Voit2018ApJ}. We find that the simulations that form multiphase gas are concentrated to the bottom right part of the figure, where either $\sigma_s$ is large or $t_\mathrm{ti}/t_\mathrm{ff}$ is short. This is in agreement with the findings of \cite{Choudhury2019}, who showed that the $\min(t_\mathrm{ti}/t_\mathrm{ff})$ required for cold gas to condense out depends on the amplitude of density fluctuations. They also showed that the $\min(t_\mathrm{ti}/t_\mathrm{ff})$ for which the gas becomes multiphase for a given $\sigma_s$ (or amplitude of density fluctuations) rises steeply once $\sigma_s\gtrsim0.5$. This effect is seen for our compressive driving run $\zeta0.0H1.0$ldens which has $t_\mathrm{ti}/t_\mathrm{ff}>10$ but still undergoes multiphase condensation. 

We attempt to construct a condensation curve like in \citet[][see their section 4]{Voit2021ApJ} with the functional form 
\begin{subequations}
\begin{equation}
    \min(t_\mathrm{ti}/t_\mathrm{ff})=\exp(c_1\sigma_s)\label{eq:condensation_curve_tti_by_tff}
\end{equation} to separate between the single phase and multiphase runs.
We choose $c_1=6$ from an empirical fit to our data. However, we have two outlier runs, $H1.0$ and its high-resolution counterpart $H1.0$HR which have $t_\mathrm{ti}/t_\mathrm{ff}\sim2$ but still do not form multiphase gas. Since this curve ignores the importance of turbulent mixing of fluctuations, it is unable to predict the occurrence of multiphase condensation correctly for runs with strong turbulent mixing. 

\subsubsection{The importance of $t_\mathrm{ti}/t_{\mathrm{mix}}$}\label{subsubsec:t_ti_tmix_criterion}
Now we discuss the effects of the ratio $t_\mathrm{ti}/t_\mathrm{mix}$ on the multiphase condensation. As discussed earlier, \cite{Gaspari2018ApJ} propose that gaseous halos become multiphase if $t_{\mathrm{ti}}/t_{\mathrm{mix}}\lesssim1$ and remain stable otherwise. This criterion does not correctly predict the outcomes of our simulations, since 7 out of the 15~halos with $t_\mathrm{ti}/t_\mathrm{mix}>1$ form multiphase gas. We think this discrepancy may partly arise because \citet{Gaspari2018ApJ} use $\delta\rho/\rho\propto\mathcal{M}$ (or $\sigma_s\propto\mathcal{M}$) to derive the amplitude of density fluctuations in their study \citep[based on the results from cluster-scale simulations in][]{gaspari2013constraining}, which would make the density fluctuations directly related to $t_\mathrm{mix}$. This is not in agreement with our results. Recent studies have shown that $\sigma_s$ depends on $\mathcal{M}$, the degree of stratification (denoted by $\mathrm{Fr}$ or $H_S$) \citep{Mohapatra2020,Mohapatra2021MNRAS} and the Mach number of the compressive component of the velocities \citep{konstandin2012,Mohapatra2022MNRASc}, which correctly predict the amplitude of $\sigma_s$ in our simulations. Thus, understanding density fluctuations in cluster environments is key to predicting the thermal stability of the halo gas. 

Similar to \cref{subsubsec:t_ti_tff_criterion}, we attempt to construct a condensation curve of the form $\min(t_\mathrm{ti}/t_\mathrm{mix})=c_2\exp(c_1\sigma_s)$. We set $c_1=6$ and $c_2=1.8$ empirically. This curve correctly predicts the outcome of simulations with $\sigma_s\gtrsim0.1$. However, this criterion ignores the importance of $t_\mathrm{ff}$. Thus it  fails to predict the outcome of the two runs with weak/no driving and strong gravity ($H1.0$wdriv and $H1.0$NoTurb) where $\min(t_\mathrm{ti}/t_\mathrm{mix})\simeq1$ but $\min(t_\mathrm{ti}/t_\mathrm{ff})$ is much larger.

\subsubsection{A new condensation curve}\label{subsubsec:new_condensation_curve}
Instead of using the two ratios $t_\mathrm{ti}/t_\mathrm{ff}$ and $t_\mathrm{ti}/t_\mathrm{mix}$ separately, we construct a new ratio $t_\mathrm{ti}/\min(t_\mathrm{mix},t_\mathrm{ff})$ by taking the minimum of the two timescales in the denominator. Our new condensation curve is given by:
\begin{equation}
    \min\left(\frac{t_\mathrm{ti}}{\min(t_\mathrm{mix},t_\mathrm{ff})}\right)=c_2\times\exp(c_1\sigma_s),\label{eq:min_tti_min_tmix_tff}
\end{equation}
where $c_1=6$ and $c_2=1.8$ are empirically determined from fitting our data.
As discussed in earlier works and in previous sections of this study, multiphase condensation is inhibited when either of these timescales are short enough. We plot the minimum value of this new ratio against $\sigma_s$ in the third row of \cref{fig:min_tti_tff_vs_sigs}. This new condensation curve clearly separates all the simulations into subsets of single phase (unshaded region) and multiphase (grey shaded region). In the limit of weakly-forced turbulence with a long $t_\mathrm{mix}$, multiphase condensation is predicted well by the $t_{\mathrm{ti}}/t_{\mathrm{ff}}$ ratio. Similarly in the limit of weak stratification, the ratio $t_{\mathrm{ti}}/t_{\mathrm{mix}}$ predicts whether multiphase condensation occurs. Our new combined criterion covers both of these cases.

Although the behaviour of the condensation curve in our study is similar to that of \cite{Choudhury2019} ($t_\mathrm{mix} \gg t_\mathrm{ff}$ in their study), we find that our curve flattens to a smaller threshold $\min(t_\mathrm{ti}/t_\mathrm{ff})$ in the limit $\sigma_s\rightarrow0$. We think this difference arises because they plot $\min(t_\mathrm{ti}/t_\mathrm{ff})$ and density fluctuations $\delta\rho$ at $t=0$ in their condensation curve, whereas we show these values just before multiphase condensation occurs. We expect $\delta\rho$ to grow (for e.g., see $H1.0$wdriv run in the fourth panel of fig.~\ref{fig:time-evolution-diffmach}) and $\min(t_\mathrm{ti}/t_\mathrm{ff})$ to decrease by $t=t_\mathrm{mp}$, which would make the results consistent with each other.

\paragraph{Predictability of the outcome of a simulation:} Here we discuss whether one can predict the occurrence of multiphase condensation for a given set of simulation parameters -- namely $\mathrm{Fr}$, $\mathcal{M}$, $\mathcal{M}_\mathrm{comp}$, and the ratio of pressure and entropy scale-heights $R_{PS}$. 
The dashed lines in the second, third and fourth rows of \cref{fig:min_tti_tff_vs_sigs} show the co-evolution of the corresponding ratios and $\sigma_s$. Except for the $H4.0$sdriv run, these ratios do not show significant variation with time (after turbulence reaches a roughly steady state). Hence, if one can determine the value of $\sigma_s$ using the simulation parameters, then one can predict whether multiphase condensation occurs. We find two expressions for $\sigma_s^2$ in the literature relevant to the turbulence parameters in our simulations:
\begin{equation}
\sigma_s^2=\ln\left(1+0.33^2\mathcal{M}^4+\frac{0.1\mathcal{M}^2R_{\mathrm{PS}}}{\left(\mathrm{Fr}+0.25/\sqrt{\mathrm{Fr}}\right)^{2}}\right),\label{eq:sigs_expression_M21}\\
\end{equation}
from \cite{Mohapatra2021MNRAS} for subsonic stratified turbulence (where $R_{PS}=H_P/H_S=0.67$ for our simulations) and 
\begin{equation}
\sigma_s^2=\ln\left(1+3\mathcal{M}_\mathrm{comp}^{1.7}\right),\label{eq:sigs_expression_K12}
\end{equation}
from \cite{konstandin2012} for compressively forced subsonic turbulence. As we show in \cref{fig:sigs_measured_vs_predicted}, \cref{eq:sigs_expression_M21} agrees well with the the measured value of $\sigma_s$ in our natural driving simulations (left column), except the `wdriv' runs. Similarly, \cref{eq:sigs_expression_K12} accurately predicts the scaling with $\mathcal{M}_\mathrm{comp}$ for our compressively driven turbulence simulations. The `wdriv' (where turbulence may not have saturated yet) and `NoTurb' runs (where we seed initial density fluctuations by hand) do not show good agreement with either scaling relation.

\begin{figure}
		\centering
	\includegraphics[width=\columnwidth]{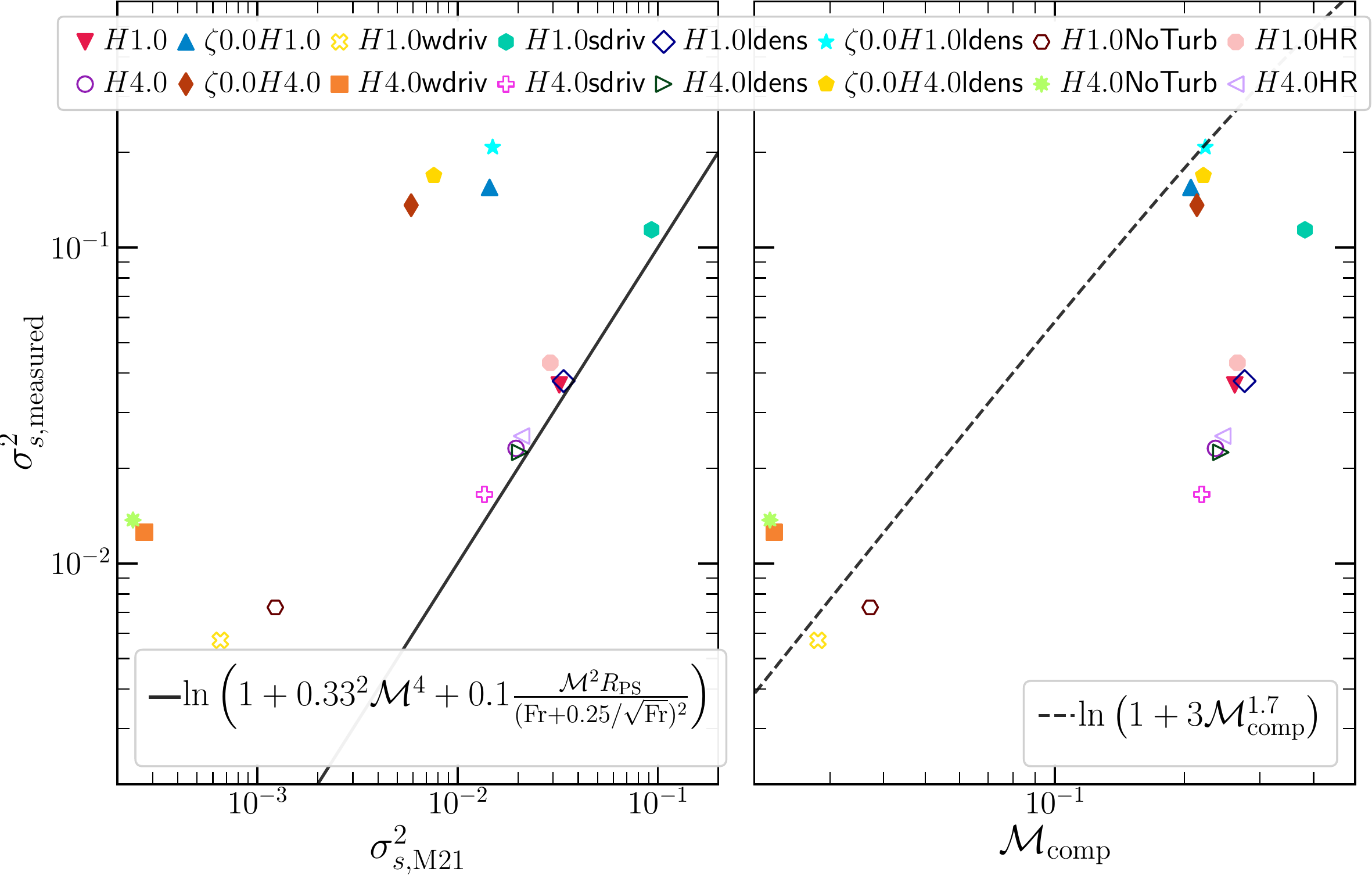}	
	\caption[dens fluc measured vs predicted]{\emph{Left column:} Scatter plot of the measured logarithmic density fluctuations squared $\sigma_{s,\mathrm{measured}}^2$ in our simulations vs.~their predicted value based on the scaling relation in \cref{eq:sigs_expression_M21}. \emph{Right column:} Scatter plot of $\sigma_{s,\mathrm{measured}}^2$ vs the compressive component of the rms Mach number $\mathcal{M}_\mathrm{comp}$. The dashed line shows the scaling relation in \cref{eq:sigs_expression_K12}. The measured $\sigma_s$ shows a remarkable agreement with \cref{eq:sigs_expression_M21} predicted values for the natural driving runs, except weak turbulent forcing (`wdriv' runs, which may not have reached a turbulent steady state yet). On the other hand, the compressive forcing ($\zeta0.0$) runs agree well with the \cref{eq:sigs_expression_K12}. The runs without driven turbulence (`NoTurb' runs) do not agree well with either of the scaling relations.}
	\label{fig:sigs_measured_vs_predicted}
\end{figure}

\paragraph{Importance of $f_\mathrm{turb}$:} Among the simulations that do not form multiphase gas, most reach a steady state where the thermal energy lost due to radiative cooling is replenished by turbulence dissipation and thermal heating. The steady state value of $\sigma_s$ varies only by a few $\%$.
However, as seen in the third row of \cref{fig:time-evolution-diffmach}, $f_\mathrm{turb}>1$ for the $H4.0\mathrm{sdriv}$ run. Thus, the heating rate due to turbulence exceeds the net cooling rate (thermal heating is switched off to prevent further over-heating). Initially, the strong turbulence drives large density fluctuations and the pink dashed line initially crosses over to the multiphase side of the condensation curve (in the fourth row of \cref{fig:min_tti_tff_vs_sigs}). However, within a few $t_\mathrm{mix}$, the gas is overheated, which increases the temperature, decreases $\mathcal{M}$ and $\sigma_s$, and raises the value of $\min{t_\mathrm{ti}}$. When $f_\mathrm{turb}>1$, even when the gas properties instantaneously satisfy the condensation criterion, the gas can be heated up on timescales $t<t_\mathrm{ti}$, and multiphase condensation is prevented.

\end{subequations}
\section{Caveats and Future Work}\label{sec:caveats-future}
Here we discuss some of the shortcomings of our study and possible ways to address them. We also outline some future prospects of this work.

\paragraph*{Resolution requirements}

In this set of simulations, all our standard set of runs  use $512^3\times768$ resolution elements to resolve the domain of size $40^2\times60~\mathrm{kpc}^3$. So the minimum length that we can resolve is $\sim80~\mathrm{pc}$. In order to capture the turbulent mixing layers between the hot- and cold-phase gas, as well as to reproduce the pressure-temperature phase diagrams, one needs to resolve the cooling length $\ell_\mathrm{cool}$, which is orders of magnitude below our resolution limit.  In particular, the clear evidence for isochoric cooling in Figure \ref{fig:dens-temp-pdf-fid} is an indication that cold gas has collapsed to the grid scale.  At that point, the gas cannot be compressed anymore because of insufficient resolution, pressure equilibrium cannot be maintained, and the gas cools isochorically.   

Further, the small-scale turbulence is also not well-resolved in this study. Hence we have not analysed the scale-by-scale kinematics of the hot and cold phases here and leave it to a follow-up study. 

We conduct two high-resolution simulations -- $H1.0$HR and $H4.0$HR with $1024^2\times1536$ resolution elements. We present these in \cref{app:convergence_test}. The results of the higher resolution simulations are similar to those presented in the main text.   However, our resolution is still far from what is required to resolve the cooling length $\ell_\mathrm{cool}$, so although the convergence in \cref{app:convergence_test} is encouraging it is far from a guarantee that the results would be the same if our resolution were sufficient to resolve all the key length-scales in the problem.

\paragraph*{Turbulence driving and heating model}
Throughout the duration of the simulation, we constantly force turbulence on large scales. Further, to prevent the model from undergoing a global runaway cooling flow, we have applied a shell-by-shell energy balance at all times. Instead of such a fine-tuned balance at all times, clusters are rather expected to undergo cycles of heating and cooling, where a cooling episode triggers strong feedback, heats the gas and prevents it from further cooling \citep[as seen in simulations, such as][]{prasad2015,Beckmann2019A&A}. In a future study, we plan to explore the effect of episodic turbulence driving and decay, to mimic AGN on-off scenarios.

\paragraph*{Missing physics}
The density-dependent heating model that we use in our simulations (defined in \cref{subsubsec:shell_balance}) is quite idealised.  We have ignored other possible heating sources such as cosmic rays \citep{Butsky2020ApJ,Su2020MNRAS,Kempski2020MNRAS}, thermal conduction \citep{Bruggen2016ApJ,Jennings2023MNRAS}, mixing of hot bubbles with the surrounding ICM \citep{banerjee2014turbulence,hillel2017}, etc. We have also ignored the effect of magnetic fields in this study. \cite{Ji2018MNRAS}  have shown that magnetic fields, independent of orientation can destabilise buoyant oscillations and modify both the amplitude and morphology of density fluctuations, which are critical to understanding the onset of multiphase condensation. \cite{Wang2021MNRAS,Mohapatra2022VSF} show that magnetic fields can modify the kinematics of both the hot and cold phases. 
We plan to conduct follow-up studies exploring the effects of some of these physical elements.

\paragraph*{Geometry}
We have modelled the ICM as a plane-parallel atmosphere with constant acceleration due to gravity. However, cluster atmospheres are expected to be spherical/elliptical. \cite{choudhury2016} showed that the amount of cold gas condensing depends on the variation of $\bm{g}$ (or $t_{\mathrm{cool}}/t_{\mathrm{ff}}$) along the radial separation from the cluster centre. 
The energy and mass budgets are also expected to be different in a spherical atmosphere, since the denser central gas has a smaller mass fraction. The hot gas would be able to expand and cool more easily compared to the plane-parallel atmosphere.
We plan to look into the effects of the cluster geometry in a future study. 

\section{Concluding remarks}\label{sec:Conclusion}
In this work, we have explored the conditions that lead to cold gas condensation from the thermally unstable hot phase in the intracluster medium. We have conducted 16 idealised simulations of a local box of size $(40^2\times60)~\mathrm{kpc}^3$ including radiative cooling, density-dependent thermal heating and turbulent driving (in 14 out of 16 simulations). 
The important time scales that govern multiphase condensation in such a system are:(1) thermal instability time $t_\mathrm{ti}(\propto t_{\mathrm{cool}}$, the cooling time); (2) gravitational free-fall time ($t_\mathrm{ff}$); and (3) turbulent mixing time ($t_\mathrm{mix}$).  A short $t_{\mathrm{ti}}$ makes condensation more likely, whereas shorter $t_\mathrm{ff}$ and $t_\mathrm{mix}$ are expected to prevent condensation. Since $t_\mathrm{cool}\propto\rho^{-1}$ (gas density), the amplitude of logarithmic density fluctuations $\sigma_s$ is also an important parameter to determine local variations in $t_\mathrm{ti}$. 
The ratios between the aforementioned timescales of the system---$t_{\mathrm{ti}}/t_{\mathrm{ff}}$ and $t_{\mathrm{ti}}/t_{\mathrm{mix}}$ are important to predict the occurrence of multiphase condensation. Here we summarise the main takeaway points of this work, focusing on the importance of these ratios:

\begin{itemize}
    \item In the limit of weak stratification, the ratio $t_{\mathrm{ti}}/t_{\mathrm{mix}}$ predicts the occurrence of multiphase condensation. We find that turbulent mixing suppresses multiphase gas condensation even for runs with $\min(t_{\mathrm{ti}}/t_{\mathrm{ff}})\simeq2$ (see $H4.0$ run in Figs.~\ref{fig:time-evolution-fid} and \ref{fig:timescales-fid}). This result is further corroborated by our findings in our strong turbulent driving set of runs (labelled `sdriv', see Figs.~\ref{fig:time-evolution-diffmach} and \ref{fig:timescales-diffmach}).  

    \item In our weak turbulence driving simulations (labelled `wdriv') and simulations without constantly driven turbulence (labelled `NoTurb' ), we find the occurrence of multiphase condensation is predicted well by the $t_{\mathrm{ti}}/t_{\mathrm{ff}}$ ratio (see Figs.~\ref{fig:time-evolution-diffmach} and \ref{fig:timescales-diffmach}). Strong stratification suppresses multiphase condensation even when $\min(t_\mathrm{ti}/t_\mathrm{mix})\simeq1$ in our $H1.0$wdriv and $H1.0$NoTurb runs. 
    
    \item Large density fluctuations always increase the likelihood of multiphase condensation. Cold gas forms in our simulations with $\min(t_{\mathrm{ti}}/t_{\mathrm{mix}})\gtrsim1$ and $\min(t_{\mathrm{ti}}/t_{\mathrm{ff}})\gtrsim10$, if the turbulence driving promotes strong density fluctuations, such as for compressive driving (see $\zeta0.0$ runs in Figs.~\ref{fig:time-evolution-fid}, \ref{fig:timescales-fid},  \ref{fig:time-evolution-lowdens} and \ref{fig:timescales-lowdens}). This happens due to the formation of dense pockets of cold gas with short $t_{\mathrm{ti}}$. The dependence of multiphase condensation on $\sigma_s$ is clearly seen in \cref{fig:min_tti_tff_vs_sigs}.
    
    \item Thus the two ratios $\min(t_{\mathrm{ti}}/t_{\mathrm{ff}})$ and $\min(t_{\mathrm{ti}}/t_{\mathrm{mix}})$ collectively predict whether multiphase condensation occurs. In the limit that one of these ratios is much larger than the other, the larger of the two determines whether multiphase gas forms. Taking into account our findings above, we propose a new condensation criterion that considers the importance of both $t_\mathrm{ff}$ and $t_\mathrm{mix}$ as well as the variability in $t_\mathrm{ti}$ due to large density fluctuations, which we parameterise using $\sigma_s$. Our new multiphase condensation criterion is given by $\min(t_{\mathrm{ti}}/\min(t_{\mathrm{mix}},t_\mathrm{ff}))=c_2\times\exp(c_1\sigma_s)$ with $c_1=6$ and $c_2=1.8$, empirically determined and shown in the bottom panel of Fig.~\ref{fig:min_tti_tff_vs_sigs}. When the minimum value of the ratio $t_{\mathrm{ti}}/\min(t_{\mathrm{mix}},t_\mathrm{ff})$ falls below this threshold, multiphase condensation occurs in our simulations. 

    \item Unlike previous studies, we find that the entropy scale height does not always play a significant role in determining whether or not a system forms multiphase gas. Turbulent mixing flattens the entropy gradient on scales smaller than the driving scale in a few mixing time-scales. However, in the limit of weak or no turbulence, simulations with a steeper entropy gradient are more stable against thermal condensation.

    \item Our simulations that form multiphase gas reach a second steady state after most of the condensed cold gas rains down through the bottom $z$-boundary. In this state, we find the value of $\min(t_\mathrm{ti}/\min(t_\mathrm{mix},t_\mathrm{ff}))$ to be independent of the initial value of $\min(t_\mathrm{ti}/(t_\mathrm{mix},t_\mathrm{ff}))$ (before the condensation begins). Instead, its steady state value increases with the amplitude of turbulent density fluctuations.

\end{itemize}

\section*{Acknowledgements} 
PS acknowledges a Swarnajayanti Fellowship (DST/SJF/PSA-03/2016-17) and a National Supercomputing Mission (NSM) grant from the Department of Science and Technology, India.
CF acknowledges funding provided by the Australian Research Council (Future Fellowship FT180100495 and Discovery Projects DP230102280), and the Australia-Germany Joint Research Cooperation Scheme (UA-DAAD).
This work was supported in part by a Simons Investigator award from the Simons Foundaton (EQ) and by NSF grant AST-2107872.
We further acknowledge high-performance computing resources provided by the Leibniz Rechenzentrum and the Gauss Centre for Supercomputing (grants~pr32lo, pr48pi and GCS Large-scale project~10391), the Australian National Computational Infrastructure (grant~ek9) and the Pawsey Supercomputing Centre (project~pawsey0810) in the framework of the National Computational Merit Allocation Scheme and the ANU Merit Allocation Scheme. The analysis presented in this article was performed in part on computational resources managed and supported by Princeton Research Computing, a consortium of groups including the Princeton Institute for Computational Science and Engineering (PICSciE) and the Office of Information Technology’s High Performance Computing Center and Visualization Laboratory at Princeton University.
The simulation software, \texttt{FLASH}, was in part developed by the Flash Centre for Computational Science at the Department of Physics and Astronomy of the University of Rochester.

This work used the following software/packages:
\texttt{FLASH} \citep{Fryxell2000,Dubey2008}, \texttt{matplotlib} \citep{Hunter4160265}, \texttt{cmasher} \citep{Ellert2020JOSS}, \texttt{scipy} \citep{Virtanen2020}, \texttt{NumPy} \citep{Harris2020}, \texttt{h5py} \citep{collette_python_hdf5_2014} and \texttt{astropy} \citep{astropy2018}.

\section{Data Availability}
All relevant data associated with this article is available upon reasonable request to the corresponding author.

\section{Additional Links}
Movies of projected density and temperature as well as time-evolution of $z$-averaged timescale profiles of different simulations are available as online supplementary material, as well as at the following links:
\begin{enumerate}
    \item \href{https://youtube.com/playlist?list=PLuaNgQ1v_KMbZmcaPOEvuM-VUW7vKYxM9}{Playlist} of fiducial runs and compressive driving ($\zeta0.0$) sets of runs;
    \item \href{https://youtube.com/playlist?list=PLuaNgQ1v_KMbQpTMc6_nu5deeclJ-LN2B}{Playlist} of weak (`wdriv') and strong (`sdriv') driving sets of runs;
    \item \href{https://youtube.com/playlist?list=PLuaNgQ1v_KMZLHRP-XwHkNWIxE-s7v6u9}{Playlist} of low density (`ldens') sets of runs;
    \item \href{https://youtube.com/playlist?list=PLuaNgQ1v_KMaDkCzwFX0TG2ytT6dK4w_o}{Playlist} of high resolution (`HR') runs.
    \item \href{https://youtube.com/playlist?list=PLuaNgQ1v_KMbycgNDwPd4RyIm4yPLIzDk}{Playlist} of runs without external driving `NoTurb'.
\end{enumerate}



\bibliographystyle{mnras}
\bibliography{refs.bib} 




\appendix
\renewcommand\thefigure{\thesection \arabic{figure}} 
\setcounter{figure}{0}   
\setcounter{table}{0}   
\section{Convergence test with resolution}\label{app:convergence_test}
Here we check the convergence of the results of our fiducial set of runs by doubling the resolution of our simulations. Similar to our fiducial set, the $H1.0$HR run becomes multiphase whereas the $H4.0$HR run remains single phase till $t=t_\mathrm{end}$.

\begin{figure}
		\centering
	\includegraphics[width=\columnwidth]{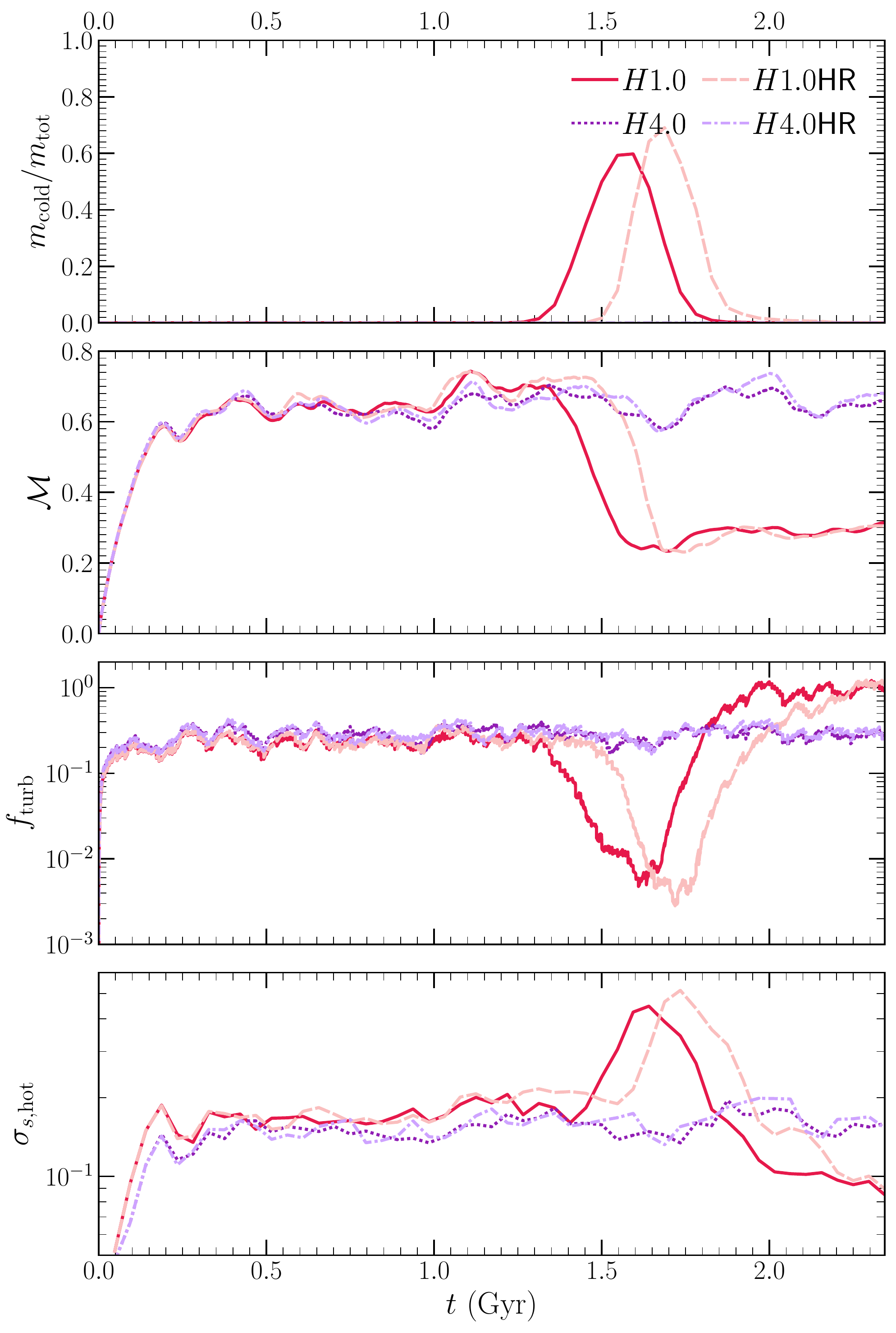}	
	\caption[Time evolution of relevant quantities for different resolution runs]{Similar to \cref{fig:time-evolution-fid}, but for our fiducial set and a higher resolution (HR) set of runs. These volume-averaged quantities are largely convergent with resolution.}
	\label{fig:time-evolution-diffres}
\end{figure}

\begin{figure*}
		\centering
	\includegraphics[width=2\columnwidth]{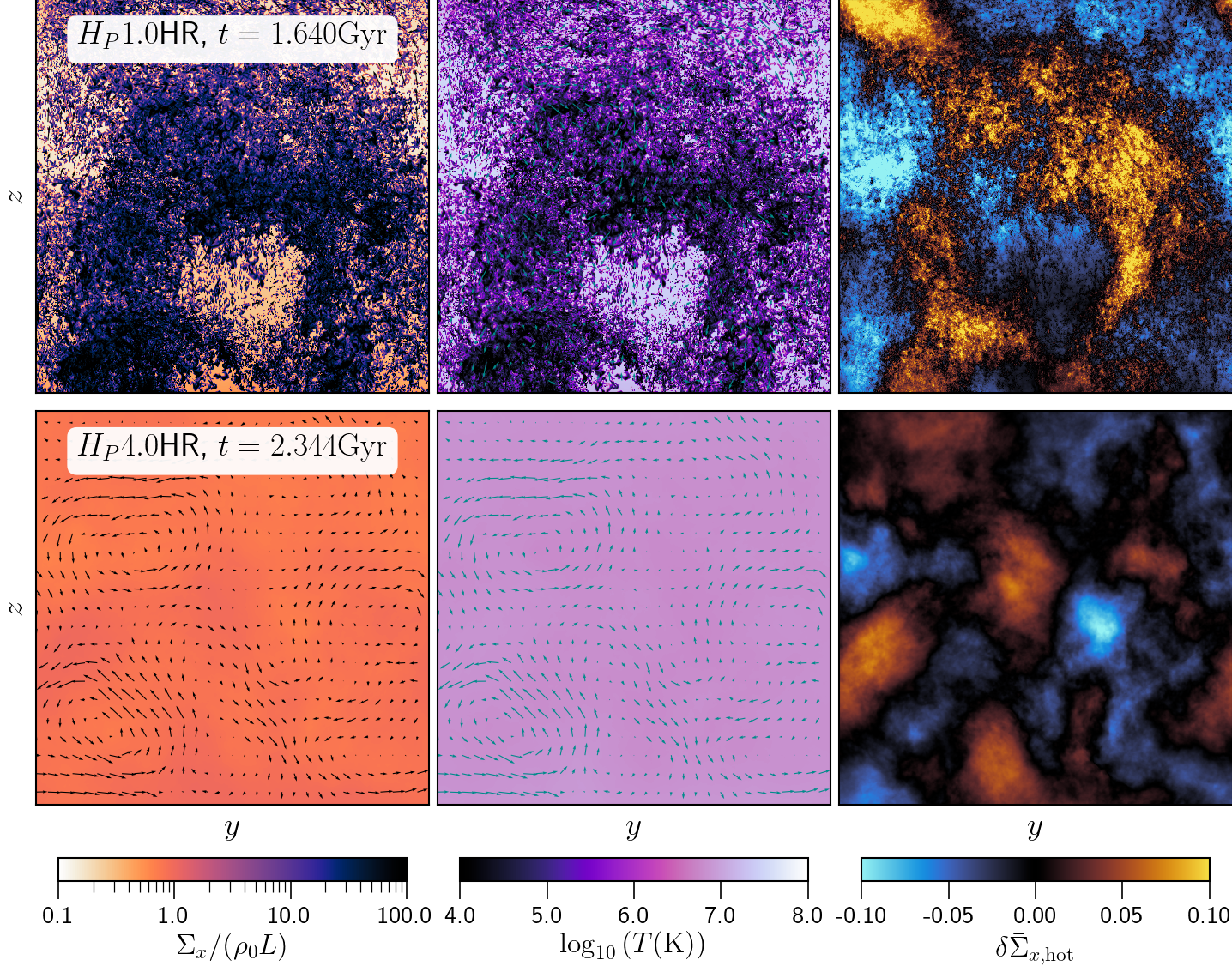}	
	\caption[Projections of density, temperature hot phase density fluctuations for fiducial runs at different resolution]{Similar to \cref{fig:projplots-fid}, but for our high-resolution set of runs. For the multiphase $H1.0$HR run, the cold gas collapses to smaller scales compared to its fiducial counterpart $H1.0$.}
	\label{fig:projplots-diffres}
\end{figure*}

We show the time-evolution of the different volume averaged quantities in \cref{fig:time-evolution-diffres}. For the single phase $H4.0$ and $H4.0$HR runs, the evolution of these quantities are quite similar and almost overlapping throughout the duration of the simulation. The $H1.0$HR run forms cold gas slightly later compared to the $H1.0$ run. However, the steady state values of all quantities before and after the formation of cold-phase gas are similar, so the results are largely in agreement. 

In \cref{fig:projplots-diffres}, we show the high-resolution counterpart of \cref{fig:projplots-fid}. Clearly, the cold gas collapses to smaller scales upon increasing resolution. We have already discussed regarding this effect in \cref{subsubsec:dens_temp_phase_diagram} and \cref{sec:caveats-future}. When the cooling length of the gas is not resolved, it collapses to the grid scale and cannot be compressed anymore.

\bsp	
\label{lastpage}
\end{document}